\documentclass{article}

\usepackage{arxiv}
\usepackage[utf8]{inputenc}
\usepackage[T1]{fontenc}
\usepackage{amsmath,amssymb,amsfonts}
\usepackage{bm}
\usepackage{graphicx}
\usepackage{array}
\usepackage{booktabs}
\usepackage{multirow}
\usepackage{tabularx}
\usepackage{longtable}
\usepackage{makecell}
\usepackage[table,xcdraw,dvipsnames]{xcolor}
\usepackage{algorithm}
\usepackage{algorithmic}
\usepackage{subfig}
\usepackage{verbatim}
\usepackage{microtype}
\usepackage{nicefrac}
\usepackage{wrapfig}
\usepackage{xspace}
\usepackage{textcomp}
\usepackage{url}
\usepackage[numbers,sort&compress]{natbib}
\usepackage[hidelinks,bookmarksopen=true,bookmarksnumbered=true]{hyperref}

\newcommand{\resultnof}[2]{#1 $\pm$ \scriptsize{#2}}
\newcommand{\ours}{\emph{QuantFormer}\xspace}
\newcommand{\qftext}[1]{#1}

\usepackage{amsmath,amsfonts,bm}

\def\eqref#1{equation~\ref{#1}}

\def\1{\bm{1}}

\DeclareMathAlphabet{\mathsfit}{\encodingdefault}{\sfdefault}{m}{sl}
\SetMathAlphabet{\mathsfit}{bold}{\encodingdefault}{\sfdefault}{bx}{n}

\def\sS{{\mathbb{S}}}

\DeclareMathOperator*{\argmin}{arg\,min}

\usepackage[utf8]{inputenc}
\usepackage{soul}
\usepackage{pifont}
\usepackage{amsmath}
\usepackage{amssymb}
\usepackage{bm}
\usepackage{color,soul}
\usepackage[dvipsnames]{xcolor}
\usepackage{gensymb}

\usepackage{xcolor}
\hypersetup{
    colorlinks,
    linkcolor={red!80!black},
    citecolor={blue!50!black},
    urlcolor={blue!80!black}
}

\newcommand{\mytilde}{\raise.17ex\hbox{$\scriptstyle\mathtt{\sim}$}}
\newcommand{\mask}{\texttt{[MASK]}\xspace}
\newcommand{\cls}{\texttt{[CLS]}\xspace}
\newcommand{\stim}{\texttt{[STIM]}\xspace}
\newcommand{\neuron}{\texttt{[NEURON]}\xspace}

\def\ee{\mathbf{e}}

\def\hh{\mathbf{h}}

\let\oldll\ll
\def\ll{\mathbf{l}}
\def\mm{\mathbf{m}}

\def\oo{\mathbf{o}}

\def\ttt{\mathbf{t}}

\def\xx{\mathbf{x}}

\def\zz{\mathbf{z}}

\def\lL{\mathcal{L}}

\def\nN{\mathcal{N}}

\def\sS{\mathcal{S}}

\def\Re{\mathbb{R}}

\hypersetup{
  hidelinks,
  pdftitle={QuantFormer: Learning to Quantize for Neural Activity Forecasting in Mouse Visual Cortex},
  pdfauthor={Salvatore Calcagno, Isaak Kavasidis, Simone Palazzo, Marco Brondi, Luca Sita, Giacomo Turri, Daniela Giordano, Vladimir R. Kostic, Tommaso Fellin, Massimiliano Pontil, Concetto Spampinato},
  pdfsubject={Neural activity forecasting with dynamic signal quantization},
  pdfkeywords={neural activity forecasting, quantization, visual cortex, transformers, computational neuroscience}
}

\renewcommand{\headeright}{Preprint}
\renewcommand{\undertitle}{Preprint}
\renewcommand{\shorttitle}{QuantFormer}

\title{\textit{QuantFormer}: Learning to Quantize for Neural Activity Forecasting in Mouse Visual Cortex}

\author{
\textbf{Salvatore Calcagno$^{1}$ \quad Isaak Kavasidis$^{1}$ \quad Simone Palazzo$^{1}$}\\
\textbf{Marco Brondi$^{2}$ \quad Luca Sit\`a$^{2}$ \quad Giacomo Turri$^{2}$}\\
\textbf{Daniela Giordano$^{1}$ \quad Vladimir R. Kostic$^{2}$ \quad Tommaso Fellin$^{2}$}\\
\textbf{Massimiliano Pontil$^{2}$ \quad Concetto Spampinato$^{1}$}\\[0.6em]
\normalfont $^{1}$PeRCeiVe Lab -- University of Catania\\
\normalfont $^{2}$Italian Institute of Technology
}

\date{}

\begin{document}

\maketitle

\begin{abstract}
Forecasting neural activity is essential for understanding complex animal behaviors and \qftext{advancing neural forecasting for potential future applications such as closed-loop optogenetics}. Existing transformer-based methods struggle with the unique spatiotemporal sparsity and intricate dependencies of neural signals. This study introduces \ours, a transformer-based model designed to forecast neural activity from two-photon calcium imaging data from mouse visual cortex. 
Unlike conventional regression-based approaches, \ours reframes the forecasting task as a classification problem via dynamic signal quantization, enabling more effective learning of sparse neural activation patterns. Furthermore, \ours tackles the challenge of analyzing multivariate signals from an arbitrary number of neurons by incorporating neuron-specific tokens, allowing scalability across diverse neuronal populations. The model is trained in a self-supervised manner using masked autoencoding with vector quantization on the Allen dataset, a large-scale mouse visual cortex benchmark. 
\qftext{\ours achieves the strongest visual-response classification results among the compared methods and performs best in capturing temporal dynamics.} Qualitative analysis also shows the effectiveness in handling sparse activations.
\ours sets a new benchmark in forecasting mouse visual cortex activity, paving the way for a foundational model in neural signal prediction. Source code is available \href{https://github.com/SalvoCalcagno/quantformer2024}{online}.
\end{abstract}

\paragraph{Impact statement.}
This work introduces a first step toward a foundational model for two-photon neural-signal forecasting without behavioral variables. Its computational efficiency may support future online neuroscience experiments.

\keywords{Neural activity forecasting \and Quantization \and Visual cortex \and Transformer models \and Interpretability \and Computational neuroscience}

\section{Introduction}
Complex animal behavior is believed to stem from the electrical activity of coordinated ensembles of neurons within specific brain circuits \cite{yuste2015neuron,yuste2024neuronal}. For example, during sensory perception~\cite{ohki2005functional,ohki2006highly} and motor coordination~\cite{harpaz2014scale,omlor2019context, santos2015corticostriatal}, correlated patterns of electrical activity in groups of neurons are observed in the primary sensory and motor cortices \cite{chen2024brain, inagaki2022neural, panzeri2022structures}. These activity patterns are structured both spatially and temporally, meaning different subsets of neurons are activated at distinct times. The patterns are further distinguished based on the sensory stimuli or motor outputs they represent \cite{kondapavulur2022transition,miller2014visual, rule2022self}. Importantly, the activity at any given moment is influenced by the recent history of the neuronal circuit \cite{boly2007baseline,leinweber2017sensorimotor,luczak2022neurons}.

Neuronal activity patterns can be recorded in the intact brain using various methods, including electrophysiological recordings \cite{jun2017fully, steinmetz2019distributed} and optical techniques such as two-photon microscopy \cite{denk1990two, helmchen2005deep} combined with fluorescent activity reporters \cite{chen2013ultrasensitive, dana2019high}. These methods enable high-resolution, in vivo imaging of brain cell activity, allowing researchers to observe coordinated neuronal responses during sensory stimulation and motor execution. For instance, studies have shown specific neuronal ensembles encoding stimulus features and behavior in the sensory cortex \cite{buetfering2022behaviorally,carrillo2019controlling}, and in the motor cortex during motor programs \cite{serradj2023task}.

A key challenge in neuroscience is developing predictive models that can forecast neuronal activity in a given brain network based on past observations. This task holds significant scientific value \qftext{and may eventually support} online closed-loop experiments, such as optogenetics, where real-time adjustments to experimental conditions can enhance intervention effectiveness. Our approach to \textbf{forecasting neural activity} differs fundamentally from traditional encoding and decoding methods. Decoding methods, such as those detailed in \cite{azabou2023POYO, Ye2023NeuralDataTransformer2, antoniades2023neuroformer}, focus on mapping internal neural variables (e.g., neural activity) to external variables, such as behavior occurring simultaneously with the neural response or the stimulus that elicited it. On the other hand, encoding methods~\cite{turishcheva2024dynamicsensoriumcompetitionpredicting, turishcheva2024retrospectivedynamicsensoriumcompetition, li2023v1t, Xu2023Multimodal, sinzStimulusDomainTransfer2018}, aim to map external variables to internal neural activity.
In contrast, our goal is to model future neural activity without relying on synchronous data, emphasizing the unique challenge of forecasting rather than decoding past stimuli/behaviors or encoding past activity.

The motivation for predicting neuronal activity stems from its demonstrated effectiveness in investigating the sensorimotor cortex of humans and nonhuman primates~\cite{truccolo2010collective}. \qftext{However, beyond its value for offline analysis, neural activity forecasting may eventually support online experimental paradigms, including closed-loop optogenetic studies.} 
A key element in achieving this goal is leveraging data that is accessible in real-time scenarios. Traditional methods often employ spiking activity data~\cite{SchrimpfKubilius2018BrainScore,PeiYe2021NeuralLatents,turishcheva2024dynamicsensoriumcompetitionpredicting}, which 
presents challenges due to limited accuracy of real-time spike inference.
\qftext{We therefore focus on baseline-normalized $\Delta{F}/F$ traces computed directly from corrected fluorescence using the pre-stimulus baseline, thereby avoiding explicit spike inference and retaining the signal in the fluorescence domain.}

In this paper, we propose \ours, a transformer-based model for two-photon calcium imaging forecasting using latent space vector quantization. Our approach reframes the forecasting problem as a classification task through vector quantization, enabling the learning of sparse activation spikes. 
Posing a regression problem as a classification task, even when the data is implicitly continuous, facilitates sparse coding (as already demonstrated in pixel \cite{oord2016pixel} and audio \cite{oord2016wavenet} spaces), which is crucial given the relative rarity of neuronal activations. 

QuantFormer first learns, in a masked autoencoding fashion~\cite{devlin2018bert,he2022masked}, to compress input neural signals into a sequence of quantized codes, thus allowing self-supervised training by predicting masked items in the sequence. This strategy facilitates the pretraining of the model for downstream tasks by quantization learning while providing a natural way to approach forecasting as the prediction of masked future codes.

Scalability to an arbitrary number of neurons is achieved by learning and prepending a set of neuron-specific tokens to the input. These tokens empower the model to process data from all available neurons without the need to create one model for each neuron or to increase its complexity through multivariate analysis, thus facilitating the effective learning of neural dynamics. 

\ours was extensively evaluated on the publicly available Allen dataset \cite{de2020large}, the only existing benchmark - to our knowledge - providing raw fluorescence traces, \qftext{and achieves the strongest forecasting results among the compared methods under the accumulated-gradient evaluation of normalized temporal dynamics.}
Ablation studies also confirm the design choices underlying \ours. 

In summary, the key contributions of this work include:
\begin{itemize}%
\qftext{\item \textbf{Single-trial neural-response forecasting}: We propose an approach that forecasts future neural responses from pre-stimulus activity and stimulus information. Such forecasts may provide a computational component for future online experimental paradigms.
}
\item \textbf{Reframing forecasting as a classification problem}: By employing vector quantization, QuantFormer learns a discrete representation of neural signals, enabling the use of classification techniques to predict sparse neuronal activations effectively.
\item \textbf{Handling arbitrary neural populations}: Our model uses neuron-specific tokens to facilitate the analysis of multivariate signals from any number of neurons, ensuring scalability and generalization across individuals and sessions.
\item \textbf{Establishing a foundation model for mouse visual cortex}: Leveraging unsupervised learning on the Allen dataset, QuantFormer demonstrates robust forecasting across different stimuli, individuals, and experimental conditions, laying the groundwork for future research in neural signal prediction.
\end{itemize}

\section{Related Work}
This paper introduces \ours, a transformer-based method trained using self-supervision for neural forecasting on two-photon calcium imaging data.\\

\begin{figure}
    \centering
    \includegraphics[width=0.95\linewidth]{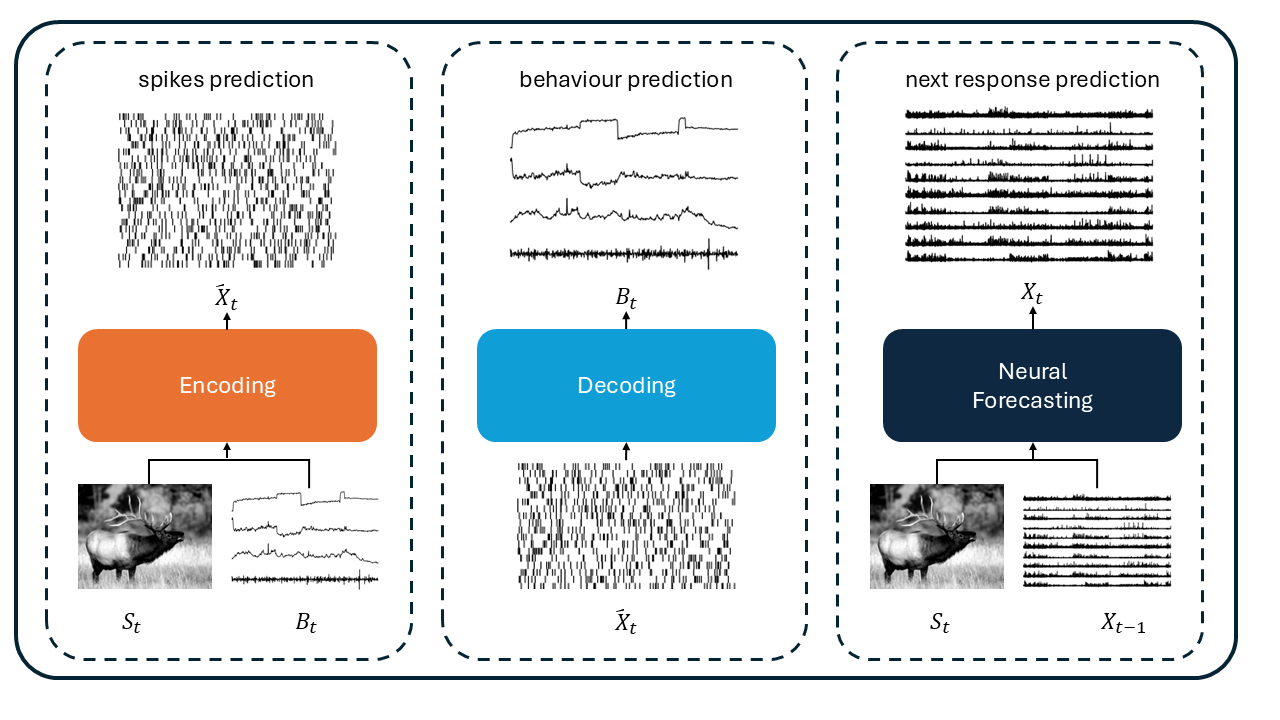}
    \caption{\textbf{Comparison of encoding, decoding, and forecasting tasks}. Encoding methods take a stimulus and behavioral variables at time $t$ to predict neural spikes at the same time point. In contrast, decoding methods do the opposite, using spike responses at time $t$ to predict behavioral variables for that time step. Neural forecasting differs from both, as it uses the stimulus at time $t$ and fluorescence traces at time $t-1$ to predict neural responses at time $t$.}
    \label{fig:tasks}
\end{figure}

In deep learning for two-photon calcium imaging, existing methods have predominantly focused on neuron segmentation \cite{soltanian2019fast, sita2022deep, bao2021segmentation, xu2023neuroseg}, as well as encoding and decoding tasks.\\
In particular, \textit{decoding methods} map neural activity (internal variables) to external outcomes like behavior \cite{azabou2023POYO, Ye2023NeuralDataTransformer2, antoniades2023neuroformer}. These methods, which use neural activity as input, focus on decoding synchronous patterns, such as behaviors occurring alongside neural responses. However, their goal is not to predict future neural dynamics but to link current neural signals to external events.\\
\textit{Encoding methods} do the opposite, mapping external variables (e.g., stimuli) to neural activity. Approaches such as \cite{turishcheva2024dynamicsensoriumcompetitionpredicting, turishcheva2024retrospectivedynamicsensoriumcompetition, li2023v1t, Xu2023Multimodal, sinzStimulusDomainTransfer2018} predict neural responses based on stimuli, but often rely on trial-averaged data and are not designed to forecast future neural activity on a single-trial basis without the use of synchronous behavior variables, which are not accessible in online settings.
\\In contrast, the task we present in this work, \textit{neural forecasting}, aims to predict future neural activity triggered by external inputs (e.g., visual stimuli) based on the neuron states, i.e., on its past neural data. \qftext{Such forecasts could inform decisions in future online experimental settings.}
The difference between encoding, decoding and neural forecasting tasks is clarified in Fig.~\ref{fig:tasks}.

In terms of model architecture, \ours is aligned with the recent trend in modeling univariate and multivariate single-dimensional time-series signals through transformers. 
LogTrans \cite{li2019logtrans} pioneered the use of transformers in univariate forecasting, utilizing causal convolutions to enhance attention locality. Informer \cite{zhou2021informer} improves efficiency in long-sequence forecasting with sparse attention. PatchTST \cite{Yuqietal2023PatchTST} handles multichannel data by processing univariate signals in patches, limiting its ability to capture inter-variable correlations. Crossformer \cite{zhang2022crossformer} applies attention across both time and variable dimensions to exploit multivariate dependencies, with cross-window self-attention capturing long-range relationships. Pyraformer \cite{liu2021pyraformer} introduces pyramidal attention to represent multiresolution features. FEDformer \cite{zhou2022fedformer} replaces self-attention with Fourier decomposition and wavelet transforms for handling seasonal data patterns.

\ours employs transformers where multivariate signals are handled through prepending tokens that encode specific neurons as well as tokens for stimulus encoding. It employs a pretraining procedure based on reconstructing masked input, in \textbf{an autoencoder configuration}, thus leveraging the extensive volumes of unlabeled neural signals from two-photon calcium imaging data in a self-supervised learning setting. 
This strategy has already demonstrated remarkable results in various research areas, including language modeling~\cite{devlin2018bert}, audio~\cite{huang2022masked}, and vision~\cite{he2022masked,tong2022videomae}. 
Additionally, during pretraining, we also learn to quantize in a manner similar to image generation~\cite{esser2021taming}. However, this quantization approach has not been applied to pose forecasting as a code classification task in the neural signal data domain.\\
Though not directly applied to two-photon calcium imaging, methodologies like BrainLM \cite{ortega2023brainlm}, based on the vanilla transformer, and SwiFT \cite{kim2023swift}, which leverages Swin transformers \cite{liu2021swin} trained on fMRI data, are more closely aligned with our approach, as they perform pretraining through self-learning. Both models are then tuned on downstream tasks, though only BrainLM includes brain state forecasting. \\

Finally, regarding the data, all the encoding and decoding methods discussed above rely either on deconvolved calcium traces or spiking data (as illustrated in Fig.~\ref{fig:tasks}). While spiking data reflects a more processed stage of neural signals, its practical application in online settings is limited due to the challenges of real-time spike inference, which often misses a significant portion of spikes \cite{spikes2021}.
Given these limitations, we opted for raw fluorescence traces, \qftext{which avoid explicit spike inference and may therefore provide a suitable signal representation for future online neural forecasting.} This decision inherently guided us towards the Allen Visual Coding dataset \cite{de2020large}, which provides a comprehensive large-scale benchmark for the mouse visual cortex, encompassing raw fluorescence traces (unlike other existing benchmarks such as BrainScore \cite{SchrimpfKubilius2018BrainScore}, Neural Latents'21 \cite{PeiYe2021NeuralLatents}, and SENSORIUM \cite{wang2023towards}), deconvolved traces and spiking activity.

\section{Method}
\subsection{Overview}

Our approach consists of two distinct training phases: \emph{pretraining} through masked autoencoding and \emph{downstream training} addressing neural activity classification and forecasting.
In the pretraining phase, we train a vector-quantized autoencoder to reconstruct a sequence of non-overlapping neuronal signal patches - following similar procedures from computer vision \cite{dosovitskiy2020image, he2022masked} - a fraction of which is replaced with a \mask token. The objective of this task is twofold: first, it encourages the model to learn an expressive and reusable feature representation of neuronal signal for downstream tasks; second, it lays the foundation for its use as a forecasting tool, by using \mask tokens as placeholders for future signal.

In the downstream phase, we employ the pretrained encoder to predict neuron activations in response to visual stimuli. As mentioned above, this prediction task can be framed as a time series forecasting task, with the objective of predicting the temporal development of a neuronal response. Alternatively, it can be seen as a classification problem, where an \emph{active} (i.e., neuron activation) or \emph{inactive} (i.e., normal neuron activity) label is associated to the neural signal recorded after stimulus visualization. 

\begin{figure*}
    \centering
    \includegraphics[width=\textwidth]{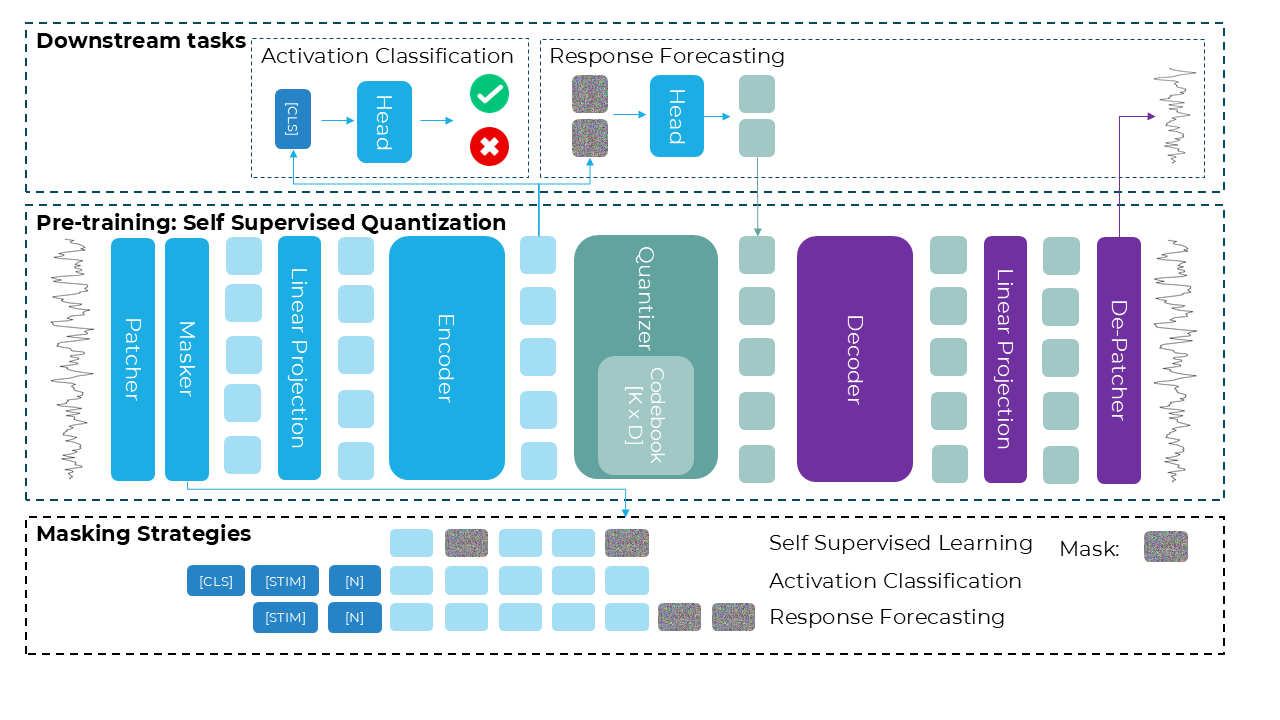}
    \caption{\textbf{\ours architecture}.
    During \textbf{pretraining} we employ a \textbf{self-supervision quantization strategy} that learns to reconstruct the randomly-masked patches along a quantization scheme. For \textbf{response forecasting}, \neuron and \stim tokens are prepended to the input, and neuronal response patches are masked; the model predicts for the masked patches quantized codes that are converted, through the quantization decoder learned during self-training, to a continuous signal. For \textbf{activation classification}, an additional \cls token is included in the sequence, and its output embedding is fed to the activation classifier.
    }
    \label{fig:architecture}
\end{figure*}

\subsection{Problem formulation}

Let $\sS = \left\{ s_1, \dots, s_S\right\}$ be the set of stimuli to which subjects can be exposed, and let $\nN = \left\{n_1, \dots, n_N\right\}$ be the set of neurons under analysis. We define an observation $\oo = \left( \xx_b, \xx_f, n, s, a \right)$ to be the set of signals associated to neuron $n \in \nN$ when presenting stimulus $s \in \sS$: $\xx_b \in \Re^{L_b}$ is the \emph{baseline activity}, i.e., the neuronal activity \emph{before} the stimulus onset, while $\xx_f \in \Re^{L_f}$ is the \emph{response activity}, i.e., the neuronal activity \emph{after} the stimulus onset; $a \in \left\{ 0, 1 \right\}$ denotes whether neuron $n$ is \emph{active} after the presentation of stimulus $s$, and $L_b$ and $L_f$ denote the temporal length of the different portions in the recorded signals, for a given sample rate $r$. According to~\cite{chen2013ultrasensitive}, a neuron is marked as \emph{active} ($a = 1$) if the response window has an average gain of 10\% over the average baseline luminescence. 

The ultimate goal of the proposed approach is to predict neuronal activity in response to a stimulus, by modeling either $p(\xx_f | \xx_b, n, s)$ (when posing the task as time series forecasting) or $p(a | \xx_b, n, s)$ (when posing the task as a classification problem).

\subsection{Pre-training stage}

We pose our self-supervised pretraining as a masked autoencoding task, with the objective of learning a general representation that models neuronal activity patterns with a view towards response forecasting. In order to make the representation as general as possible (as downstream training will be responsible for specialization), in this stage we aim to reconstruct \emph{the entire signal} for an observation $\oo$, i.e., the concatenation of $\xx_b$ and $\xx_f$, while \emph{ignoring both the neuron identity and the presented stimulus}. Formally, let $p(\xx)$ be the distribution of concatenated baseline and response neuronal signals, with $\xx \in \Re^{L_b + L_f}$, and let $p(\mm | \xx)$ be a masking function that removes a random portion from $\xx$. We want to learn a latent representation $\zz$, from which an estimate of the unmasked signal $\xx$ can be reconstructed as $p(\xx | \zz)$. Following common practices, we model $p(\zz | \mm)$ and $p(\xx | \zz)$ as an encoder-decoder network sharing the latent representation. Additionally, we introduce a quantization layer~\cite{huh2023improvedvqste} on the latent representation, in order to enforce that the latent representations are mapped to a predefined set of embeddings. While not strictly necessary for pretraining, quantization yields a twofold usefulness for our purposes. First, it enables a categorical representation of neuronal signal components, allowing downstream tasks to pose forecasting as a classification problem rather than a regression one, which has been shown to be easier to optimize~\cite{oord2016wavenet}. Second, quantization addresses the sparsity of neuronal activations as it forces the model to focus on a limited number of prototypes, encouraging reuse of codes corresponding to common patterns and potentially reducing the impact of over-represented components.
We thus define a set of embeddings $E = \left\{\ee_1, \dots, \ee_K\right\}$, with $\ee_i \in \Re^d$ and $K$ being the codebook dimension. In this setting, we distinguish between the continuous distribution $p(\zz_e | \mm)$ produced by the encoder network and the categorical distribution $p(\zz_q | \mm)$ obtained after quantization, defined as:
\begin{equation}
p(\zz_q = k | \mm) = 
\begin{cases}
1 & \text{with~} k = \argmin_i \left\| \zz_e(\mm) - \ee_i \right\|_2\\
0 & \text{otherwise}
\end{cases}
\end{equation}

The decoder is then supposed to learn the distribution of $p(\xx | \zz_q)$. In order to train the entire model, due to impossibility of backpropagating through the quantization operator, a straight-through estimator~\cite{yin2019understanding} is employed, directly copying the gradient of the reconstruction loss $\lL_\text{rec}$ with respect to the quantized representation, $\nabla_{z_q} \lL_\text{rec}$, to the output of the encoder. We implement $\lL_\text{rec}$ as the weighted mean squared error between the original unmasked signal $\xx$ and the decoder's output, separately taking into account the masked portion $\xx_m$ and the unmasked portion $\xx_u$:
\qftext{
\begin{equation}
\lL_\text{rec}(\xx,\hat{\xx}) = \alpha\lL_\text{MSE}(\xx_m,\hat{\xx}_m) + \lL_\text{MSE}(\xx_u,\hat{\xx}_u),
\label{eq:lrec}
\end{equation}
where $\alpha=2$ assigns greater weight to the reconstruction of masked elements. Class imbalance is handled through balanced-minibatch construction during pretraining. For each initially collated minibatch $\mathcal{B}_0$ of observations $\oo_j=(\xx_{b,j},\xx_{f,j},n_j,s_j,a_j)$, we define
\begin{equation}
\begin{aligned}
\mathcal{B}_{+}
&=\{\,\oo_j\in\mathcal{B}_0:a_j=1\,\},\\
\mathcal{B}_{-}
&\sim\operatorname{Subsample}\bigl(\{\,\oo_j\in\mathcal{B}_0:a_j=0\,\},
|\mathcal{B}_{+}|\bigr),
\end{aligned}
\end{equation}
and optimize the reconstruction loss over $\mathcal{B}=\mathcal{B}_{+}\cup\mathcal{B}_{-}$. Thus, all active observations in $\mathcal{B}_0$ are retained together with an equal random sample of inactive observations. This balanced-minibatch construction is possible because the model processes one neuronal trace at a time; multivariate approaches, such as Crossformer~\cite{zhang2022crossformer}, cannot balance individual active and inactive traces because each input contains multiple neurons.}
We complement the reconstruction loss with quantization and commitment losses from~\cite{huh2023improvedvqste}, in order to simultaneously train the encoder and learn the codebook.

From an implementation perspective, we employ transformer architectures to model both the encoder and the decoder. Similarly to common approaches in computer vision, we segment the input signal $\xx$ into a set of \emph{patches} $\left\{ \xx_1, \dots, \xx_P \right\}$, with $\xx_i \in \Re^{(L_b + L_f)/P}$ (padding can be applied to make the dimensionality an integer value). A linear projection transforms each patch $\xx_i$ into a \emph{token} $\ttt_i \in \Re^d$, which includes positional encoding. The masking function $\mm$ replaces a fraction $P_m$ of tokens with a learnable \mask token with the same dimensionality as each $\ttt_i$, producing a masked sequence $\mm = \left\{ \mm_1, \dots, \mm_P \right\}$. Following the above formulation:
\begin{equation}
p(\mm | \xx) = p(\mm_1, \dots, \mm_P | \ttt_1, \dots, \ttt_P) = \prod b_i , 
\end{equation}
where each $b_i$ is a Bernoulli random variable with probability $P_m$, such that $\mm_i = \text{\mask}$ if $b_i = 1$, and $\mm_i = \ttt_i$ otherwise. A masked input $\mm$ is then fed to the transformer encoder $\zz_e$ and quantized into $\zz_q$, keeping the same dimensionality as the masked input, i.e., $\zz_q \in \Re^{P \times d}$. The transformer decoder models $p(\xx | \zz_q)$ and includes a final projection layer that restores the patch dimensionality from the token representation; merging the resulting patches yields the reconstructed neuronal signal.

\subsection{Downstream tasks}
After pretraining the encoder-decoder network, we employ it for adaptation to specific downstream tasks, namely, \emph{neuron activation prediction} and \emph{response forecasting}.

\subsubsection{Neuronal activation prediction}
 
Given an observation $\oo = \left( \xx_b, \xx_f, n, s, a \right)$, our goal is to predict whether the target neuron responds to the stimulus or not, by modeling $p(a | \xx_b, n, s)$. We adapt the pretrained encoder to this task, with some modifications designed to inject neuron-specific and stimulus-specific knowledge, which was ignored during the self-supervised training. We first introduce a learnable \cls token~\cite{devlin2018bert,dosovitskiy2020image}, whose representation at the output of the encoder is fed to a linear binary classifier, marking the neuron as \emph{active} or \emph{inactive}. We then define a set of stimulus-specific learnable tokens $\left\{ \text{\stim}_1, \dots, \text{\stim}_S \right\}$, one for each possible stimulus, and a set of neuron-specific learnable tokens $\left\{ \text{\neuron}_1, \dots, \text{\neuron}_N \right\}$, one for each neuron under analysis; all $\text{\stim}_i$ and $\text{\neuron}_j$ tokens are $d$-dimensional vectors, i.e., with the same dimension as the encoder input tokens. Given the observation $\oo = \left( \xx_b, \xx_f, n, s, a \right)$, we segment and project the baseline signal $\xx_b$ into tokens $\left\{ \ttt_1, \dots, \ttt_P \right\}$, and then feed the encoder network with $\left\{ \text{\cls}, \text{\neuron}_n, \text{\stim}_s, \ttt_1, \dots, \ttt_P \right\}$. 

Feeding neuron and stimulus identifiers to the  encoder is a key aspect of the approach: since our backbone does not inherently handle multivariate data, we compensate for this lack by providing learnable neuron identifiers, making the model able to learn distinct activation patterns for each neuron; similarly, stimuli identifiers provide a means for the model to discover the specific stimuli a neuron responds to. Also, we only feed the baseline signal $\xx_b$ to the encoder, since the response $\xx_f$ likely contains neuron activation information, which would defeat the purpose of the classifier.

The transformer-based architecture also allows us to tackle this task in two different training settings: \emph{prompt-tuning} and \emph{fine-tuning}. With prompt-tuning, only soft prompts (i.e., \cls, all $\text{\stim}_i$ and all $\text{\neuron}_j$) can be optimized, while the encoder remains frozen. With fine-tuning, all encoder parameters can be \qftext{updated}. In this task, neither the quantizer nor the decoder are used.

\subsubsection{Neuronal response forecasting}

The objective of this task is to model $p(\xx_f | \xx_b, n, s)$, in order to predict the response time series $\xx_f$ from the baseline signal $\xx_b$, preceding the stimulus onset. A possible approach to this problem consists in using the pretrained encoder-decoder network, by masking all input tokens related to the portion of signals to be predicted, and read the forecast signal as the encoder output. However, as mentioned above, the pretrained model lacks neuron/stimulus specialization, which is needed to handle different neuronal activity dynamics. Moreover, while the pretrained encoder is trained to capture the underlying patterns of the input data for filling in missing information, this does not necessarily imply the capability to directly predict future values of a time series. To address these issues, we act in two ways: similarly to the previous task, $\text{\stim}_i$ and $\text{\neuron}_j$ tokens are added to the beginning of the sequence, to provide the model with specific information; second, rather than fine-tuning the entire model, we append a classification network after the encoder for predicting the codebook indices corresponding to masked tokens only. This approach, besides simplifying the architecture, can be specifically tailored to understand the dynamics of neuronal activity post-stimulus.

Given an input observation $\oo = \left( \xx_b, \xx_f, n, s, a \right)$, we convert the baseline signal $\xx_b$ into tokens $\left\{ \ttt_1, \dots, \ttt_P \right\}$, and construct the encoder input as $\left\{ \text{\neuron}_n, \text{\stim}_s, \ttt_1, \dots, \ttt_P, \text{\mask}, \dots, \text{\mask} \right\}$: the number of \mask tokens, $M$, depends on the length $L_f$ of the response signal. $\text{\stim}_i$ and $\text{\neuron}_j$ tokens are learned separately from their counterparts in the activation classification downstream task. 
We denote the output of the encoder corresponding to masked tokens as $\left\{ \hh_1, \dots, \hh_M \right\}$, and feed it to a classification network $\phi$, implemented as a multi-layer perceptron. We compute the set of targets $\left\{ y_1, \dots, y_M \right\}$, with $y_i \in \left\{ 1, \dots, K \right\}$, by feeding the full signal, i.e., the concatenation of $\xx_b$ and $\xx_f$ to the original pretrained encoder, reading out the quantization \qftext{indices} into which the response portion is encoded. We then train the classifier and learn the soft prompts by optimizing the cumulative cross-entropy loss over masked tokens:
\begin{equation}
\lL_\text{rf} = - \sum_{i=1}^M \log{\phi\left( \hh_{i}\right)_{y_i} }
\end{equation}
with $\phi\left(\hh_{i}\right)_c$ being the $c$-th component of the predicted class distribution for the $i$-th masked token. At inference time, the predicted codebook \qftext{indices} replace the masked tokens and the entire sequence is fed to the pretrained decoder for reconstructing the forecast response. Note that both the codebook and the decoder are frozen at this stage, while the encoder can be frozen too (thus learning soft prompts only during training) or optionally fine-tuned. 

\section{Experimental Results}

\subsection{Dataset}
The Allen Brain Observatory Dataset comprises over 1,300 two-photon calcium imaging experiments, organized into more than 400 containers. Each container, representing all the experimental data from a single mouse, consists of three 90-minute sessions \qftext{in which multiple stimuli are administered}. 
We selected 11 containers (i.e., mice) previously used in \cite{sita2022deep}. Each container has at least three complete sessions available. The original dataset includes various types of stimuli: drifting gratings, static gratings, natural scenes, natural movies, locally sparse noise and spontaneous activity \cite{allenStimInfo}. However, we excluded natural movies, as isolating individual neuron responses is challenging, and spontaneous activity, as it is not stimulus-related. Within each session, every stimulus type is presented across three distinct subsessions. Each stimulus may be shown once or multiple times. The presentation of a single stimulus, along with its corresponding neural response, is referred to as a trial. In total, we used 236,808 multivariate signals representing neuron responses from the selected mice (additional details in Table A-1 in the appendix).\\
Moreover, in our classification downstream task, we examine whether there is a response to a given stimulus within a defined response window. Across all mice and their neurons, we identify a total of 2,287,735 positive (\emph{active}) responses, while normal activity (non-responsive, \emph{inactive}) samples amount to approximately 40 million.\\
\qftext{To mitigate temporal correlations and prevent overlap between training and test sets, we partition the data on a per-subsession basis. Specifically, we allocate two subsessions for training and one for testing, with each subsession separated by 10-15 minutes. Furthermore, we ensure that training and testing data are distinctly separated by exposing the mouse to other stimuli during the interim period, thus eliminating potential temporal correlations between signals. The \qftext{exact} timestamps used for splits are reported in Section B of Appendix}.

\subsection{Training Procedure and Metrics}
\ours is pretrained in self-supervision as a masked autoencoding task through quantization, using data from all subjects.
The full model consists of 6 layers and 8 attention heads for both the encoder and the decoder, with a hidden size $d$ of 128 and a mask ratio $P_m$ of 0.2. 
We train with Adam \cite{kingma2014adam} for 50 epochs, a learning rate of $10^{-4}$ and a batch size of 32. In both downstream tasks, we fine-tune the encoder for 100 epochs with a learning rate of $10^{-3}$. The length of baseline and response signals in each observation is respectively 3 seconds and 2 seconds at sample rate $r = 30$, resulting in padded sequence lengths $L_b = 96$ and $L_f = 64$.
The number of quantized codes $K$ is set to 32.
As we diverge from these values we note that performance decreases in both tasks (see Tables A-3 and A-4 in the \emph{Appendix}). This confirms the sparsity of crucial information in brain signals, which can be encoded with as few as 32 indices (the performance decrease was less sensitive to the dimensionality $d$).
Downstream tasks were conducted separately for each subject and stimulus category, \qftext{with results reported as mean and standard deviation across subject–stimulus combinations}.  We also evaluate generalization using the \emph{leave-one-category-out} strategy, excluding specific stimulus categories or mice from pretraining and using the excluded data for downstream training.
\qftext{As metrics, we use balanced accuracy, precision, recall, and $F_1$ for classification, and MSE, SMAPE, Pearson correlation, and structural similarity index (SSIM) for forecasting. The main forecasting results use accumulated-gradient normalization, while complementary evaluations using training-set per-neuron $z$-score and median/IQR normalization, together with peak-timing and peak-amplitude metrics, are reported in Tables A-10 and A-11 of the \emph{Appendix}.}
The selected competitors for our approach, based on code availability and adaptability to the tasks, are Autoformer~\cite{autoformer}, Informer~\cite{zhou2021informer}, Crossformer~\cite{zhang2022crossformer}, and BrainLM~\cite{ortega2023brainlm}. We use BrainLM pretrained on large fMRI data (due to observed similarities between mice and humans~\cite{Eppig2015}), fine-tuned on our data (BrainLM$_{ft}$), and trained from scratch. Additionally, we include a simple LSTM-based baseline, that we empirically found to mostly predict the signal's mean\qftext{, together with two causal convolutional forecasting baselines, TCN~\cite{bai2018tcn} and Seq-U-Net~\cite{stoller2020sequnet}}. All experiments are conducted on a workstation with an 8-core CPU, 64GB RAM, and an NVIDIA A6000 GPU (48GB VRAM). 
\qftext{Detailed model-specific training configurations, parameter counts, pretraining strategies, and neural-data adaptations are reported in the \emph{Appendix}, together with the forecasting results for TCN and Seq-U-Net.}

\subsection{Results}
We initially focus on assessing model performance in stimulus-response classification; results are shown in Table~\ref{tab:classification}.
\begin{table*}[!ht]
    \centering
    \renewcommand{\arraystretch}{1.2}
        \caption{\textbf{Performance on stimulus-response classification}. All metrics marked with * have $p \oldll 0.01$, while metrics with ** have $p<0.05$ using one-sided Wilcoxon test. \qftext{Values are reported as mean $\pm$ standard deviation across container--stimulus pairs}.}
    \footnotesize{
    \begin{tabular}{l|cccc}
        \toprule
        \textbf{Method} & \textbf{Acc} ($\uparrow$)& \textbf{$F_1$} ($\uparrow$)& \textbf{Prec} ($\uparrow$)& \textbf{Rec} ($\uparrow$)\\
        \midrule
        \rowcolor{gray!10}
        LSTM & \resultnof{61.53}{12.75}*  & \resultnof{31.00}{27.71}*  & \resultnof{40.07}{39.42}*  & \resultnof{35.08}{22.91}*  \\
        Autoformer & \resultnof{58.50}{06.11}*  & \resultnof{26.21}{17.06}*  & \resultnof{65.67}{27.18}*  & \resultnof{17.36}{12.47}*  \\
        \rowcolor{gray!10}
        Informer & \resultnof{60.77}{07.13}*  & \resultnof{28.69}{15.53}*  & \resultnof{55.59}{24.82}*  & \resultnof{23.11}{15.69}*  \\
        BrainLM & \resultnof{59.66}{13.97}*  & \resultnof{22.71}{32.79}*  & \resultnof{27.25}{39.39}*  & \resultnof{19.56}{02.83}*  \\
        \rowcolor{gray!10}
        BrainLM$_{ft}$ & \resultnof{62.31}{14.67}*  & \resultnof{29.33}{03.48}*  & \resultnof{36.58}{42.11}*  & \resultnof{24.91}{29.69}*  \\ 
        Crossformer & \resultnof{75.51}{04.45}**  & \resultnof{63.89}{07.59}**  & \resultnof{85.71}{03.73}  & \resultnof{51.49}{09.03}**  \\ 
        \midrule
    \rowcolor{gray!10}
        \textbf{QuantFormer} & \resultnof{\textbf{77.39}}{03.88}  & \resultnof{\textbf{66.94}}{06.51}  & \resultnof{\textbf{85.89}}{00.04}  & \resultnof{\textbf{55.27}}{07.82}  \\  
        \bottomrule
    \end{tabular}}
    \label{tab:classification}
\end{table*} 
\begin{table*}[htb!]
    \centering
    \renewcommand{\arraystretch}{1.2}
    \caption{\textbf{Performance on stimulus-response forecasting} of \ours compared to existing forecasting methods. All metrics marked with * have $p \oldll 0.01$, while metrics with ** have $p<0.05$ using one-sided Wilcoxon test. \qftext{Values are reported as mean $\pm$ standard deviation across container--stimulus pairs.}}
    \footnotesize{
    \begin{tabular}{l|cccc}
        \toprule
        \textbf{Method} & \textbf{MSE} ($\downarrow$) & \textbf{SMAPE} ($\downarrow$) & \textbf{Corr} ($\uparrow$) & \textbf{SSIM} ($\uparrow$)\\
        \midrule
        \rowcolor{gray!10}
       LSTM & \resultnof{60,349.339}{-}* &\resultnof{0.943}{0.037}* &\resultnof{0.273}{0.129}* & \resultnof{0.003}{0.004}* \\
        Autoformer & \resultnof{19.828}{11.728}* &\resultnof{0.800}{0.033}* & \resultnof{0.312}{0.085}* & \resultnof{0.011}{0.005}* \\
        \rowcolor{gray!10}
        Informer& \resultnof{0.285}{0.376}** & \resultnof{0.707}{0.051}* & \resultnof{0.302}{0.033}* & \resultnof{0.022}{0.017}* \\
        BrainLM & \resultnof{0.605}{2.852} & \resultnof{0.701}{0.158}* & \resultnof{0.253}{0.125}* & \resultnof{0.008}{0.034}*\\
        \rowcolor{gray!10}
        BrainLM$_{ft}$ & \resultnof{0.457}{0.825} &\resultnof{0.697}{0.183}*  & \resultnof{0.337}{0.112}** & \resultnof{0.001}{0.035}*\\
        Crossformer & \resultnof{2.011}{2.749}* &  \resultnof{0.723}{0.062}* & \resultnof{0.292}{0.087}* & \resultnof{0.036}{0.020}*\\
        \midrule
        \rowcolor{gray!10}
        \textbf{QuantFormer} & \resultnof{\textbf{0.247}}{0.078}& \resultnof{\textbf{0.656}}{0.137}& \resultnof{\textbf{0.338}}{0.075}& \resultnof{\textbf{0.069}}{0.062}\\
        \bottomrule
    \end{tabular}}
    \label{tab:forecasting}
\end{table*}

\begin{figure*}
    \centering
    \includegraphics[width=0.45\textwidth]{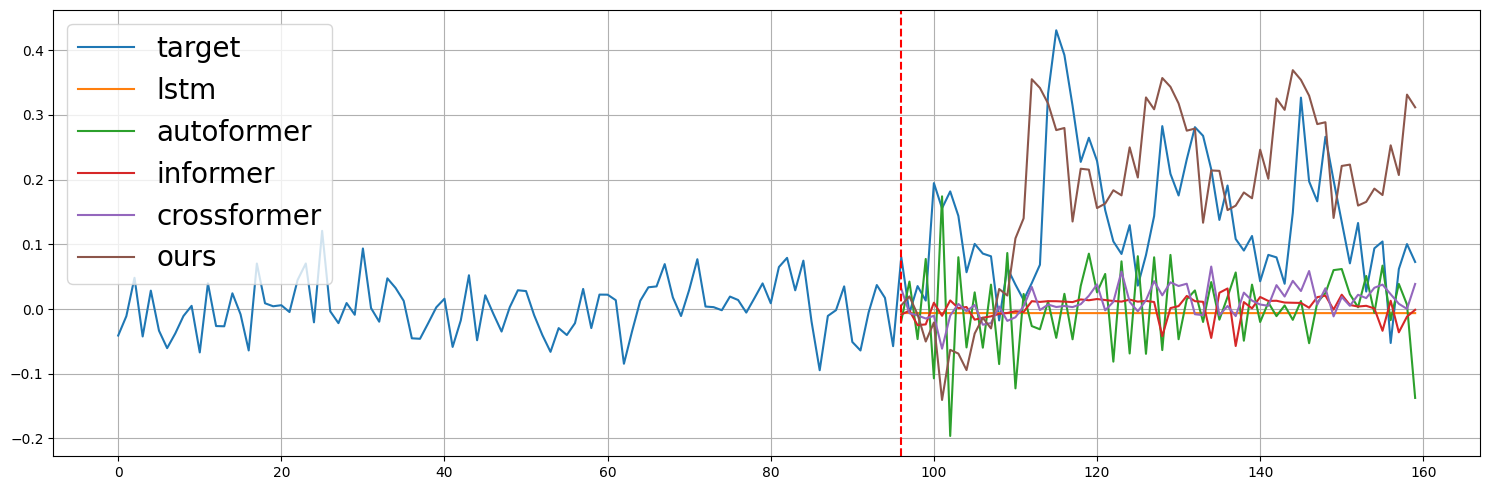}
    \includegraphics[width=0.45\textwidth]{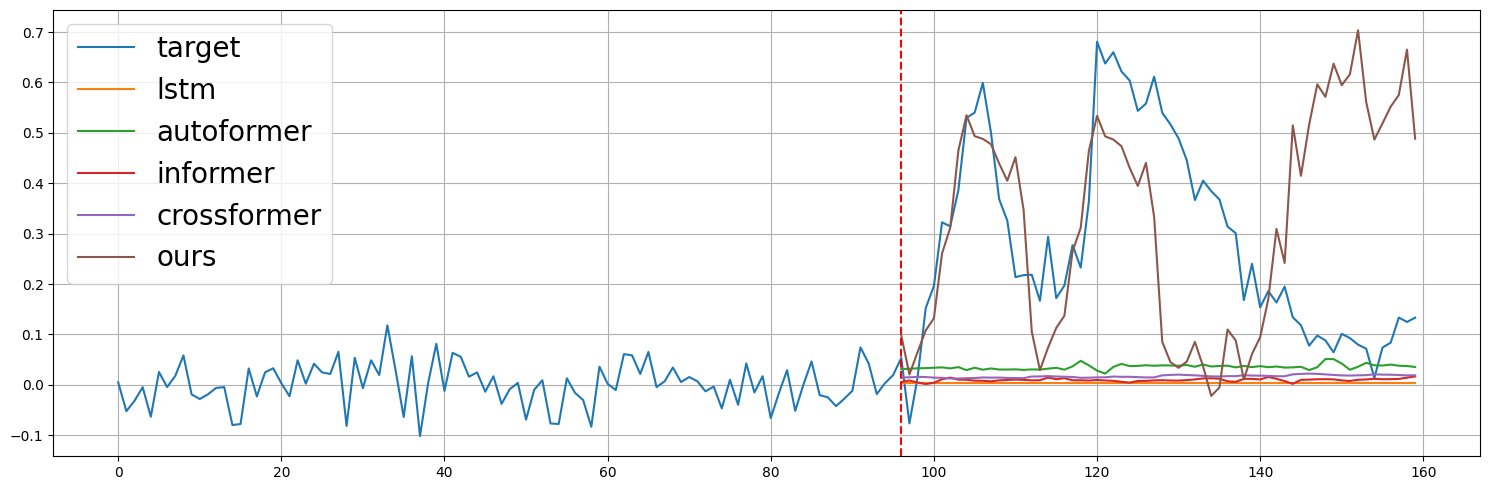}\\
    \includegraphics[width=0.45\textwidth]{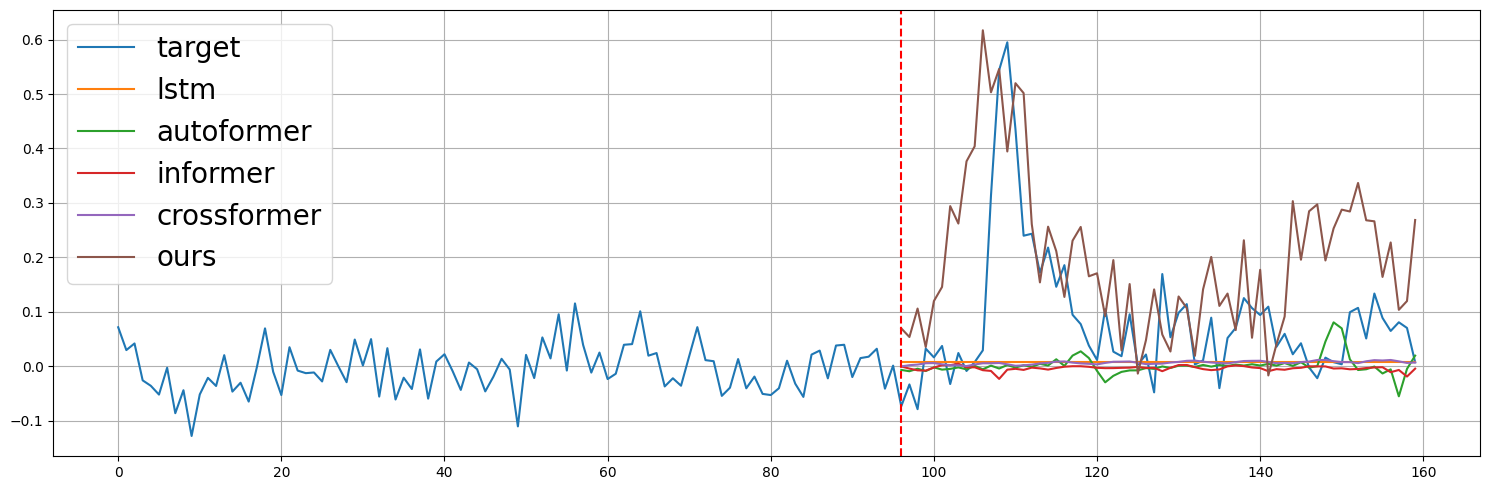}
    \includegraphics[width=0.45\textwidth]{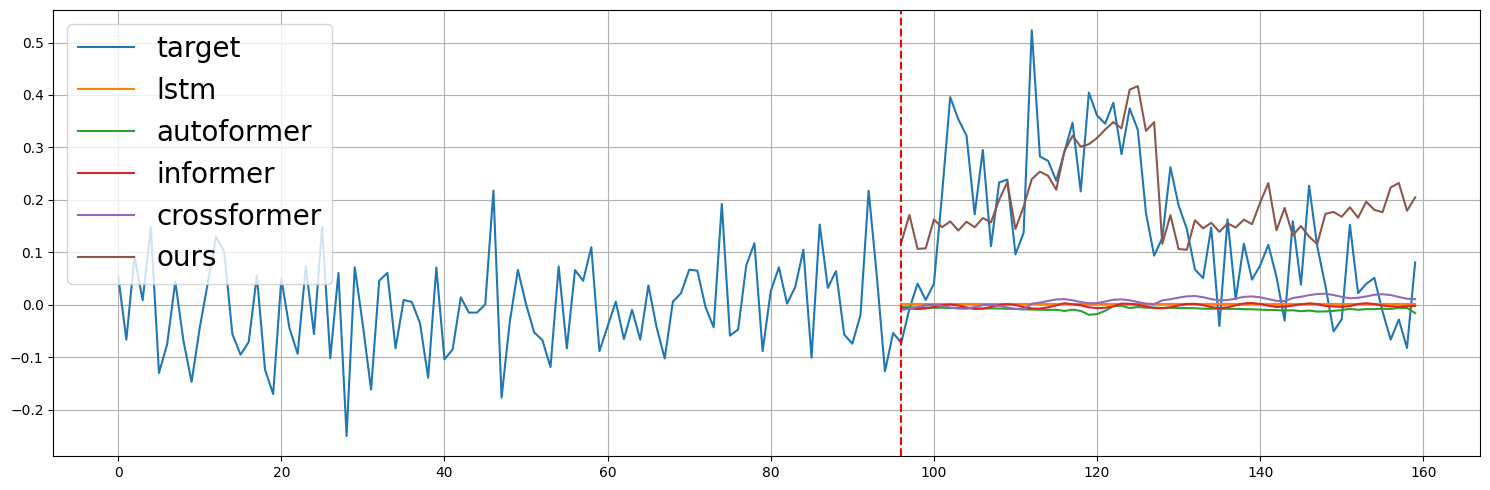}
    \caption{\textbf{Qualitative analysis of stimuli response forecasting performance by \ours and its competitors}: forecasting examples for each type of stimuli: drifting gratings (top-left), static gratings (top-right), natural scenes (bottom-left) and locally sparse noise (bottom-right). More examples can be found in Section I of the \emph{Appendix}.}
    \label{fig:forecasting}
\end{figure*}

\qftext{
Table~\ref{tab:forecasting} reports forecasting performance under accumulated-gradient normalization, in which each target and prediction is independently scaled by its accumulated temporal variation (see Section H of the \emph{Appendix}). Under this specific evaluation protocol, \ours obtains the lowest mean MSE and SMAPE and the highest mean correlation and SSIM among the compared methods. These scores characterize agreement in normalized temporal dynamics and should not be interpreted as measures of absolute forecasting accuracy on baseline-normalized $\Delta F/F$ responses. Complementary prediction-independent evaluations using training-derived per-neuron $z$-score and median/IQR normalization are reported in Tables A-10 and A-11 of the \emph{Appendix}, together with peak-timing and peak-amplitude metrics. Taken together, these evaluations distinguish point-wise reconstruction accuracy, agreement in normalized temporal dynamics, and preservation of stimulus-evoked transient timing and amplitude. Figure~\ref{fig:forecasting} provides qualitative examples of the fluorescence responses forecast by \ours and the competing methods.
}

Cross-referencing classification (Table~\ref{tab:classification}) and forecasting performance (Table~\ref{tab:forecasting}), it becomes apparent that \qftext{\ours performs strongly in both tasks}, unlike other methods such as Informer~\cite{zhou2021informer} and Crossformer~\cite{zhang2022crossformer}, which specialize in only one. For instance, while Informer exhibits good forecasting metrics, its classification metrics, especially recall, fall short. This may stem from Informer generating responses with activations surpassing the mean signal, but not reaching the threshold for positive classification. Conversely, Crossformer achieves good classification accuracy but \qftext{obtains less consistent normalized-dynamics forecasting results}, likely due to its tendency to predict constant responses that lead to positive classifications while diverging from actual response patterns.\\

To substantiate the design choices behind \emph{QuantFormer}, we conduct an ablation study to analyze the importance of different components in the model architectures for classification and forecasting tasks, focusing only on ``drifting gratings'' stimuli for simplicity. 
We start by evaluating the performance of our encoder backbone when trained from scratch, using cross-entropy for classification and MSE for forecasting (referred to as $Baseline$ in Table~\ref{tab:ablation}). The model is provided with pre-stimulus neuronal activity together with the [STIM] token. 
We then extend this by prepending the sequence with the [NEURON] token (indicated as $Learnable$ $tokens$ in Table~\ref{tab:ablation}).
Additionally, we evaluate the effects of quantization pretraining on model performance compared to pretraining using a standard autoencoder scheme without quantization (indicated as $AE$ in Table~\ref{tab:ablation})\\

\begin{table*}[htb!]
    \centering
    \renewcommand{\arraystretch}{1.2}
        \caption{\textbf{Ablation study for learnable tokens and quantization on ``drifting gratings'' stimuli}. \qftext{Values are reported as mean $\pm$ standard deviation across containers.} %
        }
    \footnotesize{
    \begin{tabular}{ll|cc|cc}
        \toprule
        & &\multicolumn{2}{c}{\textbf{Classification}}&\multicolumn{2}{c}{\textbf{Forecasting}}\\
        \midrule
        & \textbf{Method} & \textbf{Acc} ($\uparrow$)& \textbf{$F_1$} ($\uparrow$)& \textbf{MSE} ($\downarrow$)& \textbf{Corr} ($\uparrow$)\\
        \midrule
        \rowcolor{gray!10}
        & Baseline &\resultnof{{75.88}}{4.08} & \resultnof{{64.32}}{6.62}  & \resultnof{{0.021}}{0.015}  & \resultnof{{0.147}}{0.077} \\ 
        & \hspace{0.1 cm} {$\hookrightarrow$Learnable tokens} &\resultnof{{77.53}}{3.89} & \resultnof{{67.29}}{6.39} & \resultnof{0.023}{0.018}  & \resultnof{0.147}{0.055} \\ 
        \rowcolor{gray!10}
        & \hspace{0.3 cm} {$\hookrightarrow$AE} & \resultnof{{77.22}}{4.25}  & \resultnof{{66.10}}{7.41}  & \resultnof{0.019}{0.014} & \resultnof{0.207}{0.082} \\     & \hspace{0.3 cm} {$\hookrightarrow$Fixed-Bins Quantization} & \resultnof{{76.46}}{4.14}  & \resultnof{{66.15}}{7.26}  & \resultnof{0.031}{0.03} & \resultnof{0.165}{0.085} \\     
        \midrule
        \rowcolor{gray!10}
        & \hspace{0.3 cm}
        {$\hookrightarrow$Quantization} & \resultnof{{\textbf{77.66}}}{3.78}  & \resultnof{{\textbf{67.42}}}{6.35}  & \resultnof{\textbf{0.016}}{0.009}  & \resultnof{\textbf{0.252}}{0.095}\\               
        \bottomrule
    \end{tabular}}
    \label{tab:ablation}
\end{table*}
We also explore pretraining benefits for Informer~\cite{zhou2021informer}, Autoformer~\cite{autoformer}, and Crossformer~\cite{zhang2022crossformer} using quantization and standard autoencoding. However, their architectures face two challenges (details in Sect. J
of the \emph{Appendix}): 1) combining channel and time information in embeddings creates an information bottleneck, making quantization impractical for temporal patterns; 2) the imbalance between sparse activations and normal signals requires a training strategy targeting individual neurons. Unlike methods that process all neurons simultaneously, \ours uses \neuron tokens to capture individual neuron dynamics.\\
\qftext{To further isolate the effect of learned discretization, we added a fixed equal-width binning baseline with the same number of discrete indices as QuantFormer ($K=32$).}
\qftext{The fixed-bin baseline remains clearly below learned quantization, suggesting that the gain is not only due to discretization, but also to learning data-adaptive codebook prototypes.}
\begin{table*}[htb!]
    \centering
    \renewcommand{\arraystretch}{1.2}
        \caption{\textbf{Generalization performance of \ours across subjects and stimuli.} \qftext{Values are reported as mean $\pm$ standard deviation across the held-out subjects, sessions, or stimulus categories used in each leave-one-out setting}. %
        }
    \footnotesize{
    \begin{tabular}{ll|cc|cc}
        \toprule
        & &\multicolumn{2}{c}{\textbf{Classification}}&\multicolumn{2}{c}{\textbf{Forecasting}}\\
        \midrule
        & & \textbf{Acc} ($\uparrow$)& \textbf{$F_1$} ($\uparrow$)& \textbf{MSE} ($\downarrow$)& \textbf{Corr} ($\uparrow$)\\
        \midrule
        \rowcolor{gray!10}
        & Subjects & \resultnof{{77.32}}{4.04} & \resultnof{{67.18}}{6.58}  & \resultnof{{0.367}}{0.558} & \resultnof{0.344}{0.154} \\ 
        & Sessions & \resultnof{77.11}{4.08} & \resultnof{64.63}{6.82}  & \resultnof{0.615}{0.535} & \resultnof{0.438}{0.156} \\ 
        \rowcolor{gray!10}
        & Stimuli & \resultnof{{76.78}}{3.88} & \resultnof{{67.45}}{6.26}  & \resultnof{{0.411}}{0.578} & \resultnof{{0.392}}{0.142} \\ 
        \bottomrule
        
    \end{tabular}}
    \label{tab:generalization}
\end{table*}
We then assess the generalization performance of \ours on different subjects, \qftext{sessions} and stimuli with a leave-one-out strategy. Table~\ref{tab:generalization} shows that \ours generalizes effectively across various scenarios, with performance metrics similar to those in Table~\ref{tab:forecasting}, underscoring its potential as a foundational model for large-scale studies of the mouse visual cortex.

\subsection{Interpretability Analysis}
\begin{figure*}[!htb]
    \centering
    \includegraphics[width=0.45\textwidth]{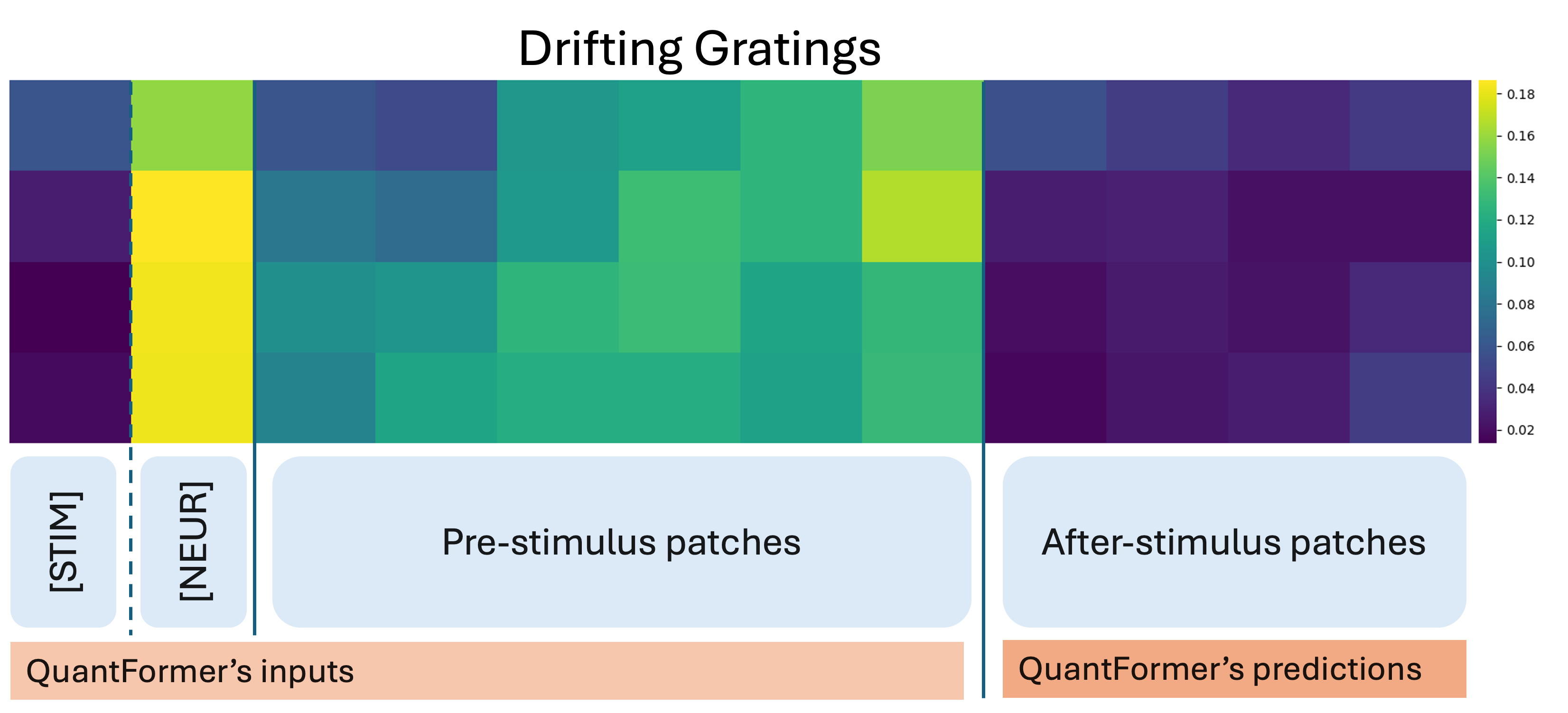}
    \includegraphics[width=0.45\textwidth]{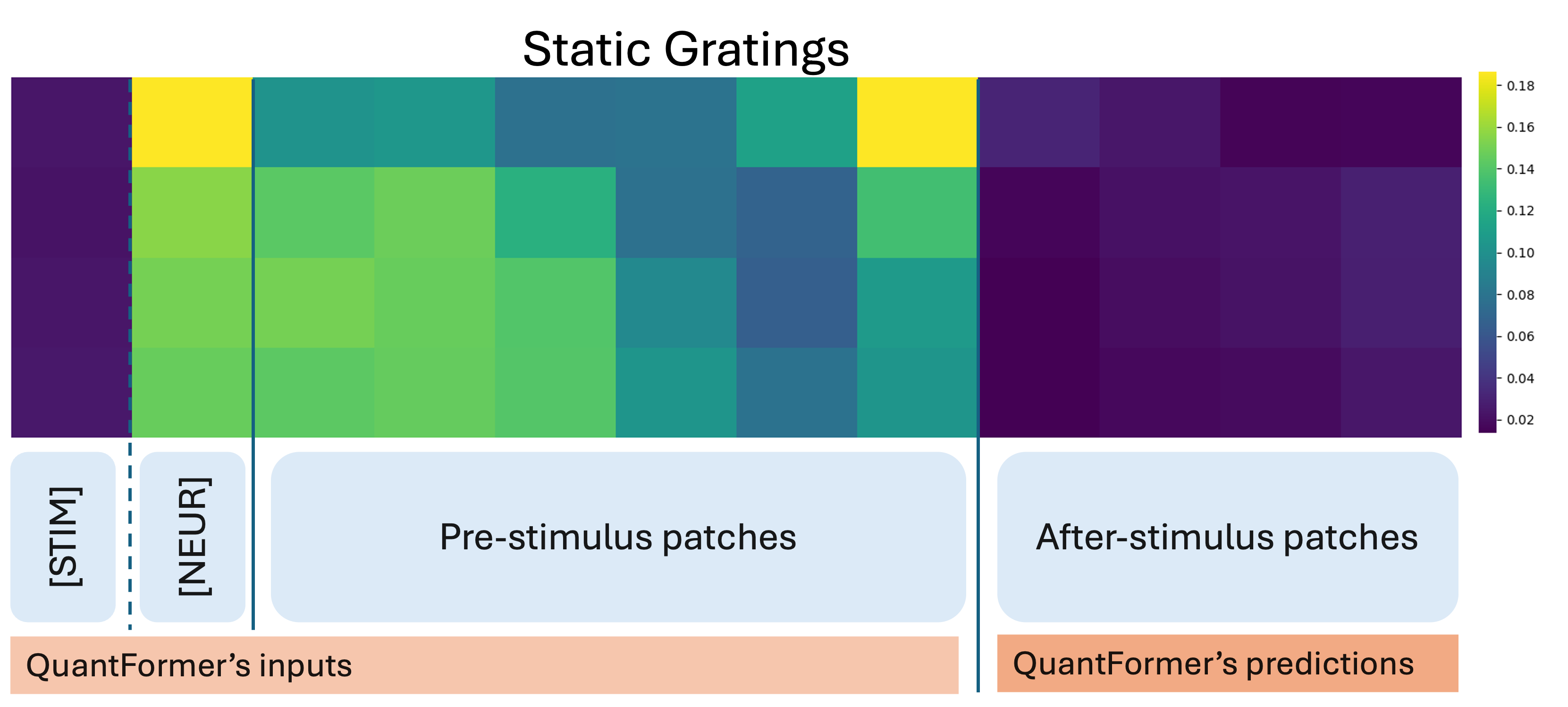}
    \includegraphics[width=0.45\textwidth]{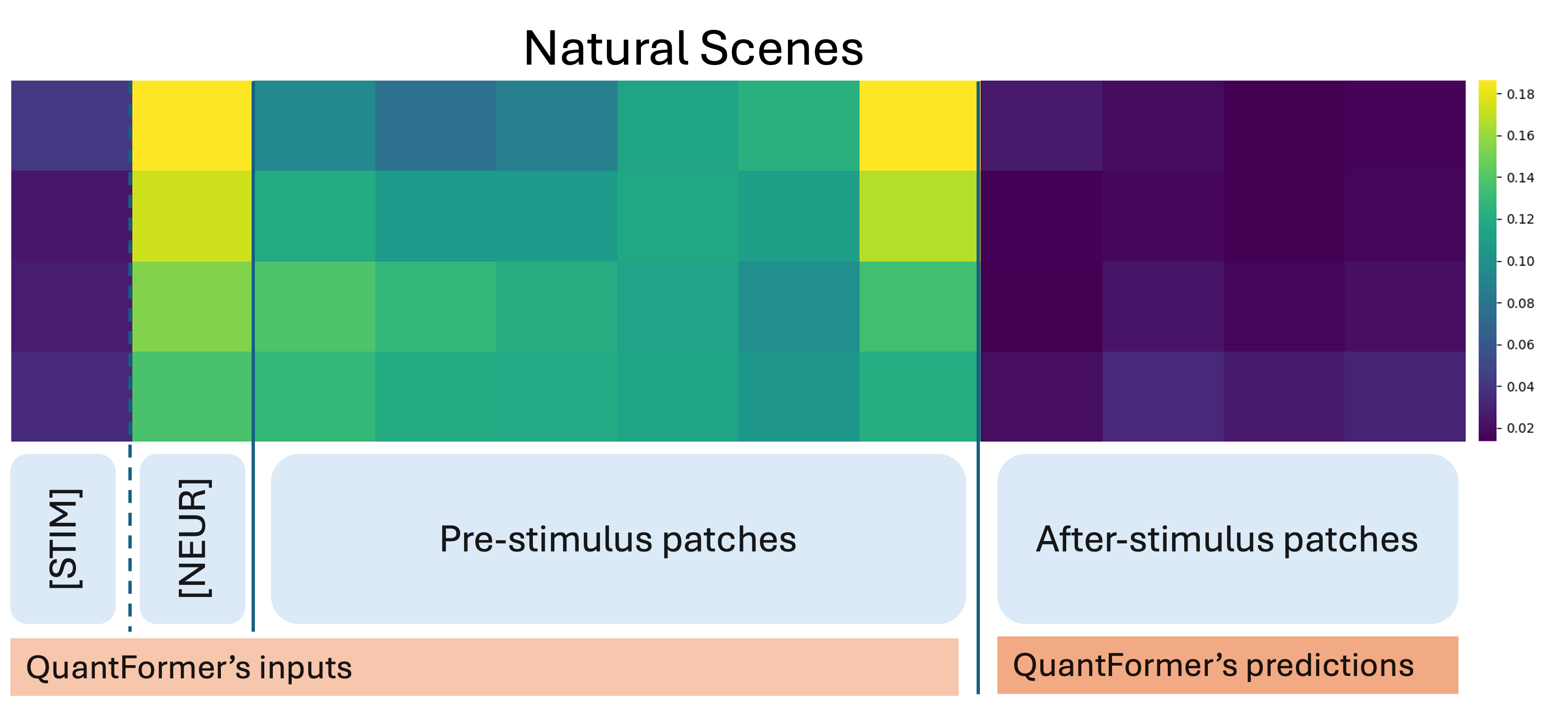}
    \includegraphics[width=0.45\textwidth]{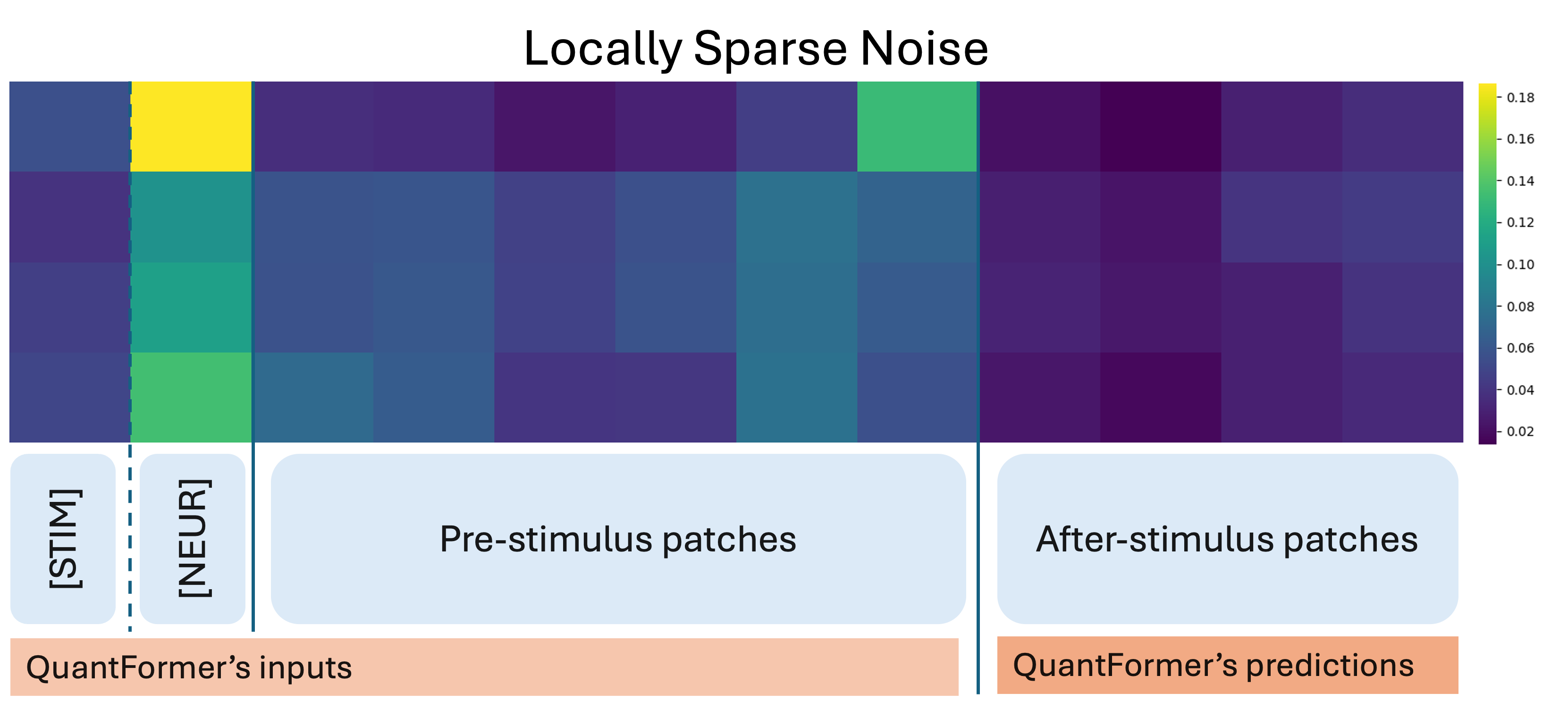}
    \caption{Attention maps for all stimulus types. Each row corresponds to a predicted code, while columns represent the $[STIM]$ token, $[NEURON]$ token, pre-stimulus patches, and the predicted codes. Color intensity indicates attention strength, with yellow denoting the most attended tokens and darker shades indicating less attended ones. This visualization highlights the model's selective focus across different components of the input and predicted outputs.}
    \label{fig:all_attention}
\end{figure*}

As an additional analysis, we examined attention score maps and the latent space of discrete codes and neuron embeddings to understand activation predictions and model interpretability. 

\textbf{Attention maps}.
We present in Fig.~\ref{fig:all_attention} attention score maps computed through attention-rollout  on \ours for neuron activation prediction for all the four types of stimuli: drifting gratings, static gratings, natural scenes, and locally-sparse noise. These maps reveal that \neuron token activity predominantly influences predictions, followed by pre-stimulus patches and stimulus token, with the model adapting pre-stimulus information based on the specific stimuli delivered. These attention maps show distinct activation patterns requiring further investigation by neuroscientists.

\textbf{Interpretability of learned codes and neuron embeddings}.
To observe which patterns are encoded in the latent space of the vector quantizer, we performed 2D t-SNE on a learned codebook. We found that variations in amplitude and patterns are captured within the latent space structure.
Additionally, we observed that the reconstructed sequence is not merely a simple concatenation of patterns; rather, the shapes and intensities of each reconstructed code strongly depend on the surrounding sequence. Specifically, we fixed a code—often associated with a peak—and varied the surrounding codes with bursts of other codes. We found that the amplitude and shape of the fixed code changed based on the surrounding context. This indicates that while codes represent patterns, the reconstruction process is influenced by the overall sequence structure. Visual examples of these findings are provided in Appendix L, Fig. A-8. 
To understand what is encoded into neuron embeddings, we also visualized through t-SNE neuron embeddings from a downstream task. We find that neuron embeddings encode information such as activation frequency and response statistics. Some examples of t-SNE neuron embeddings are shown in Appendix M, Fig. A-9.

\section{Conclusion}
We presented \emph{QuantFormer}, a transformer-based model using latent space vector quantization to capture sparse neural activity patterns in two-photon calcium imaging. By framing the regression problem as classification and leveraging unsupervised vector quantization, \qftext{\ours achieves the strongest results among the compared methods in response classification and in the accumulated-gradient evaluation of normalized temporal dynamics. The complementary prediction-independent evaluations further characterize point-wise reconstruction accuracy and highlight \ours's advantages in the timing and amplitude of stimulus-evoked transients.} Trained and tested on a subset of the Allen dataset, it excels in learning sparse activation spikes and capturing long-term dependencies, making it a versatile and robust tool for understanding neural dynamics.\\
\qftext{One} limitation of \emph{QuantFormer} \qftext{is} the lack of an inhibition mechanism, \qftext{which} may lead to sequences \textbf{with sustained high activation}, contrary to the typical single activation observed in biological neurons. \qftext{Moreover, a post-hoc deconvolution analysis (details in Section K of Appendix) shows that neither \ours nor the competing forecasting methods reliably reconstruct thresholded spike events from predicted fluorescence traces, indicating that the current formulation should be interpreted as a fluorescence forecasting method rather than a spike/event forecasting model.}
As future work, \ours will be trained on the entire Allen dataset, as well as adapted to spiking neural data (in order to use other existing benchmarks), to enhance generalization capability for creating a foundation model for the mouse visual cortex. \qftext{Although \ours{}’s non-autoregressive inference and measured latency are compatible with online processing, closed-loop control, robustness to experimental noise and drift, and the impact of forecasting errors on intervention decisions remain to be investigated.}

\section*{Acknowledgment}
S. Calcagno, I. Kavasidis, and C. Spampinato acknowledge financial support from PNRR MUR project PE0000013--FAIR.

\clearpage
\appendix
\section{Supplementary Material}
\begingroup
\raggedbottom

\setcounter{table}{0}
\setcounter{figure}{0}
\renewcommand{\thetable}{A-\arabic{table}}
\renewcommand*{\thesubsection}{\Roman{subsection}}
\renewcommand{\thefigure}{A-\arabic{figure}}

\subsection{Dataset detailed information}
The Allen Brain Data Observatory is a resource from the Allen Institute for Brain Science that provides a comprehensive collection of data on the mouse visual cortex. This resource is designed to facilitate research and understanding of brain function, particularly in the context of how sensory information is processed.  It contains various types of data regarding the mouse visual cortex ranging from cell connectivity to spontaneous neuronal activity and to stimulus-response data. 

For the experiments conducted in this work, as explained in the dataset description subsection, we used the responses to the following four types of stimuli: 

\begin{itemize}
\item \textbf{Drifting gratings:} A full field drifting sinusoidal grating at a spatial frequency of 0.04 cycles/degree was presented at 8 different directions (from 0° to 315°, separated by 45°), at 5 temporal frequencies (1, 2, 4, 8, 15 Hz). Each pattern was shown for 2 seconds, followed by 1 second of a uniform gray background before the next pattern appeared. Also blank sweeps (shown every 20 gratings) are included in this type of stimulus. Each condition (combination of temporal frequency and direction) was presented 15 times across session A. The response time was evaluated on a window of 2 seconds after the stimulus onset.

\item \textbf{Static gratings:} A full field static sinusoidal grating was presented at 6 different orientations (separated by 30°), 5 spatial frequencies (0.02, 0.04, 0.08, 0.16, 0.32 cycles/degree), and 4 phases (0, 0.25, 0.5, 0.75). Each stimulus was presented for 0.25 seconds, without intergray period. \qftext{Blank sweeps, shown every 25 gratings, are also included.} Each condition (combination of spatial frequency, orientation and phase) was presented 50 times across session B. The response time was evaluated on a window of 0.5 seconds after the stimulus onset.

\item \textbf{Locally Sparse Noise:} This type of stimulus consisted of a 16 x 28 array of pixels, each 4.65 degrees on a side. In each medium gray frame of the stimulus (presented for 0.25 seconds) a small number (11) of pixels were randomly changed to be white or black. 9,000 different frames was presented once across session C. The response time was evaluated on a window of 0.5 seconds after the stimulus onset.

\item \textbf{Natural Scenes:} 118 natural images selected from Berkeley Segmentation Dataset (Martin et al., 2001), van Hateren Natural Image Dataset (van Hateren and van der Schaaf, 1998), McGill Calibrated Colour Image Database (Olmos and Kingdom, 2004) were presented in grayscale for 0.25 seconds each, with no inter-image gray period. Each image was presented ~50 times, in random order, and the response period was \qftext{evaluated over a 0.5-second window after} the stimulus onset. 
\end{itemize}
The experimental \qftext{setup} is depicted in Fig.~\ref{fig:stim}.

\begin{figure}[h!]
\begin{center}
\includegraphics[width=0.45\textwidth]{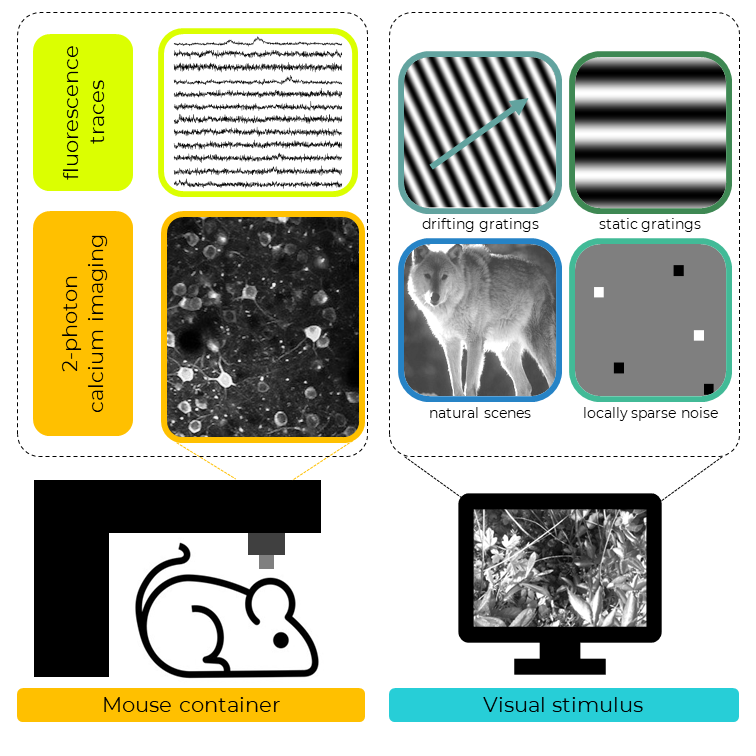}
\caption{The Allen dataset. Fluorescence time series are extracted from the two-photon calcium images (\textit{Left}). Examples of the stimuli used (\textit{Right}).}
\label{fig:stim}
\end{center}
\end{figure}

\begin{table*}%
    \centering
        \caption{\textbf{Stimuli administration protocol, dataset information and experiment durations.} Window refers to the length of data (seconds after the administration of the corresponding stimulus) considered for response forecasting. Duration is the time in minutes needed for executing a downstream training epoch for the corresponding stimulus type. The average number of neurons per mouse is 241, with a standard deviation of 63.}
    \footnotesize {
    \begin{tabular}{l|ccc|cc|c}
        \toprule
        
        & \multicolumn{3}{c}{\textbf{Acquisition protocol}} & \multicolumn{2}{c}{\textbf{Dataset information}} & \multicolumn{1}{c}{\textbf{Duration}}\\
       \midrule
       \textbf{Stimuli type} & \textbf{\# instances} & \textbf{\# trials}  & \textbf{window (s)} &  \textbf{\# mice} & \textbf{\# signals} &  \textbf{\# time (m)}\\
        \midrule

        Drifting gratings & 41 & 628  &  2 & 11 & 6,908& 0.43\\ 
        \rowcolor{gray!10}
        Static gratings & 120  & 6,000& 0.5 & 11 & 66,000& 2.4\\ 
        Locally sparse noise  & 9,000& 9,000& 0.5 & 11 & 99,000& 2.04\\
        \rowcolor{gray!10}
         Natural scenes& 118  & 5,900& 0.5 & 11 & 64,900& 1.2\\ 
        \midrule
        Total & 9,279& 21,528& -- & 11 & 236,808& 6.07\\
        \bottomrule
    \end{tabular}}
    \label{tab:dataset}
\end{table*}

The 11 container ids used for the experiments in this work are: 511507650, 511510667, 511510675, 511510699, 511510718, 511510779, 511510855, 511510989, 526481129, 536323956 and 543677425.

\qftext{\subsection{Temporal splits and reproducibility details}}

\qftext{To ensure strict reproducibility of the experimental protocol, we report in Table~\ref{tab:splits} the exact temporal train/test splits used in all experiments. For each stimulus type, the partition is defined at the subsession level: two temporally separated subsessions are used for training, while the remaining held-out subsession is used for testing. Unless otherwise specified, all experiments use random seed 42.}

\qftext{This temporal partition prevents leakage across adjacent trials, since training and test samples are drawn from distinct subsessions separated in time and interleaved with other stimulus presentations. The standard deviations reported in the result tables are computed across container--stimulus experiments, not across different random seeds.}

\begin{table*}[htbp]
\centering
\scriptsize
\renewcommand{\arraystretch}{1.1}
\caption{\qftext{\textbf{Subsession-scheme splits.}
Subsessions are Allen stimulus epoch frame ranges. Splits are defined at the subsession level: two subsessions are used for training and one held-out subsession is used for testing.}}
\label{tab:splits}

\begin{tabularx}{\textwidth}{llr l X rr}
\toprule
\textbf{Container} & \textbf{Stim.} & \textbf{Exp. ID} & \textbf{Session} & \textbf{Subsessions} & \textbf{Train trials} & \textbf{Test trials} \\
\midrule
511507650 & dg & 502115959 & A & train (746--18776, 48606--66635) test (94657--115214) & 400 & 228 \\
511507650 & sg & 501794235 & B & train (747--15198, 54931--69378) test (97370--113623) & 3840 & 2160 \\
511507650 & ns & 501794235 & B & train (16102--30550, 39580--54028) test (80213--96091) & 3812 & 2088 \\
511507650 & lns & 500855614 & C & train (737--22458, 41465--63185) test (82191--105733) & 5760 & 3120 \\
511510667 & dg & 501574836 & A & train (737--18809, 48707--66778) test (94868--115473) & 400 & 228 \\
511510667 & sg & 501498760 & B & train (737--15219, 55053--69534) test (97594--113885) & 3840 & 2160 \\
511510667 & ns & 501498760 & B & train (16126--30614, 39667--54147) test (80397--96311) & 3808 & 2092 \\
511510667 & lns & 501773889 & C & train (737--22469, 41484--63223) test (82238--105780) & 5760 & 3120 \\
511510675 & dg & 510214538 & A & train (746--18775, 48605--66634) test (94666--115223) & 400 & 228 \\
511510675 & sg & 509962140 & B & train (737--15218, 55041--69521) test (97578--113868) & 3840 & 2160 \\
511510675 & ns & 509962140 & B & train (16123--30604, 39655--54135) test (80382--96295) & 3805 & 2095 \\
511510675 & lns & 509729072 & C & train (745--22416, 41379--63051) test (82014--105490) & 5760 & 3120 \\
511510699 & dg & 502608215 & A & train (737--18809, 48710--66781) test (94869--115474) & 400 & 228 \\
511510699 & sg & 502810282 & B & train (737--15230, 55058--69540) test (97600--113892) & 3840 & 2160 \\
511510699 & ns & 502810282 & B & train (16136--30619, 39670--54152) test (80402--96317) & 3807 & 2093 \\
511510699 & lns & 502974807 & C & train (737--22463, 41473--63197) test (82212--105746) & 5760 & 3120 \\
511510718 & dg & 510514474 & A & train (746--18780, 48618--66653) test (94682--115243) & 400 & 228 \\
511510718 & sg & 510345479 & B & train (748--15198, 54943--69392) test (97388--113643) & 3840 & 2160 \\
511510718 & ns & 510345479 & B & train (16102--30558, 39589--54039) test (80229--96108) & 3808 & 2092 \\
511510718 & lns & 510174759 & C & train (746--22427, 41390--63061) test (82024--105501) & 5760 & 3120 \\
511510779 & dg & 503109347 & A & train (747--18780, 48615--66647) test (94681--115241) & 400 & 228 \\
511510779 & sg & 503019786 & B & train (746--15194, 54932--69380) test (97371--113623) & 3840 & 2160 \\
511510779 & ns & 503019786 & B & train (16098--30544, 39576--54028) test (80215--96091) & 3804 & 2096 \\
511510779 & lns & 502634578 & C & train (746--22416, 41379--63048) test (82010--105484) & 5760 & 3120 \\
511510855 & dg & 510517131 & A & train (747--18777, 48608--66638) test (94661--115220) & 400 & 228 \\
511510855 & sg & 510705057 & B & train (736--15217, 55040--69521) test (97579--113877) & 3840 & 2160 \\
511510855 & ns & 510705057 & B & train (16122--30603, 39653--54134) test (80383--96296) & 3813 & 2087 \\
511510855 & lns & 509644421 & C & train (736--22457, 41463--63192) test (82198--105729) & 5760 & 3120 \\
511510989 & dg & 501729039 & A & train (735--18806, 48716--66786) test (94873--115478) & 400 & 228 \\
511510989 & sg & 501567237 & B & train (738--15219, 55053--69534) test (97593--113884) & 3840 & 2160 \\
511510989 & ns & 501567237 & B & train (16126--30615, 39666--54147) test (80396--96311) & 3804 & 2096 \\
511510989 & lns & 501788003 & C & train (738--22467, 41481--63217) test (82229--105766) & 5760 & 3120 \\
526481129 & dg & 527048992 & A & train (746--18777, 48612--66649) test (94675--115234) & 400 & 228 \\
526481129 & sg & 528480613 & B & train (744--15194, 54930--69379) test (97381--113635) & 3840 & 2160 \\
526481129 & ns & 528480613 & B & train (16097--30547, 39578--54027) test (80217--96102) & 3807 & 2093 \\
526481129 & lns & 526768996 & C & train (744--22420, 41385--63057) test (82022--105499) & 5760 & 3120 \\
536323956 & dg & 540684467 & A & train (736--18810, 48712--66786) test (94878--115485) & 400 & 228 \\
536323956 & sg & 541048140 & B & train (735--15216, 55044--69525) test (97586--113885) & 3840 & 2160 \\
536323956 & ns & 541048140 & B & train (16122--30604, 39656--54137) test (80388--96303) & 3806 & 2094 \\
536323956 & lns & 539515366 & C & train (736--22458, 41466--63198) test (82207--105738) & 5760 & 3120 \\
543677425 & dg & 545446482 & A & train (737--18811, 48712--66786) test (94877--115484) & 400 & 228 \\
543677425 & sg & 544507627 & B & train (747--15195, 54929--69376) test (97370--113623) & 3840 & 2160 \\
543677425 & ns & 544507627 & B & train (16098--30546, 39577--54025) test (80213--96090) & 3815 & 2085 \\
543677425 & lns & 543677427 & C & train (746--22416, 41378--63048) test (82010--105486) & 5760 & 3120 \\
\bottomrule
\end{tabularx}
\end{table*}

\qftext{
\subsection{Inference-time measurements}

\qftext{To assess the computational feasibility of online inference,} we measured inference time on an NVIDIA RTX 6000 GPU using the standard forecasting configuration adopted in our experiments, with $L_b=96$ input frames and $L_f=64$ forecast frames. We evaluated a representative activity window containing 512 neurons.

A single 512-neuron activity window is processed in 6.49 ms. When processing multiple windows in parallel, throughput reaches 270.6 windows/s, corresponding to approximately $1.39 \times 10^5$ neuron-traces/s. Peak GPU memory usage is 59.5 MB for a single window and 352.8 MB at maximum throughput.
}

\qftext{These measurements characterize computational performance only and do not constitute a validation of closed-loop control or optogenetic intervention.}

\clearpage
\subsection{Hyperparameter search for quantization and embedding dimensionality}

In order to determine the optimal values for the number of quantization indices $(K)$ and embedding dimensionality ($d$), shared by both the quantized codes and the transformer models, we conduct an exploratory hyperparameter tuning on the responses to ``drifting gratings'' \qftext{and ``static gratings''} stimuli only. First, we fix the value of $d$ to 128 and perform classification and forecasting experiments varying the value of $K$. Our model achieved the best correlation score for a value of $K$ equal to 32. Afterwards, we repeated the same experiments using that number of quantized vectors and we varied the value of parameter $d$ instead.

\begin{table*}[htb!]
    \centering
    \caption{\textbf{$K$ and $d$ parameter value search for forecasting.} }

    \footnotesize {
    \begin{tabular}{c|ccccc}
        \toprule
        \multicolumn{6}{c}{\textbf{Forecasting}}\\
        \midrule
        \textbf{Value} & \textbf{MSE} ($\downarrow$)& \textbf{MAE} ($\downarrow$)& \textbf{SMAPE} ($\downarrow$)& \textbf{Corr} ($\uparrow$)& \textbf{SSIM} ($\uparrow$)\\
        \midrule
        \multicolumn{6}{c}{\textbf{Quantization indexes $K$ with $d$ = 128}}\\
        \midrule
        $K=4$  & \resultnof{0.619}{0.045} & \resultnof{0.557}{0.026} & \resultnof{0.735}{0.029} & \resultnof{0.275}{0.019} & \resultnof{-0.026}{0.023} \\
        \rowcolor{gray!10}
        $K=8$  & \resultnof{0.074}{0.012} & \resultnof{0.211}{0.012} & \resultnof{0.631}{0.013} & \resultnof{0.231}{0.014} & \resultnof{0.028}{0.008} \\
        $K=16$  & \resultnof{0.081}{0.008} & \resultnof{0.194}{0.009} & \resultnof{0.613}{0.010} & \resultnof{0.236}{0.018} & \resultnof{0.056}{0.009} \\
        \rowcolor{gray!10}
        $K=32$  & \resultnof{0.128}{0.015} & \resultnof{0.231}{0.012} & \resultnof{0.657}{0.018} & \resultnof{0.294}{0.017} & \resultnof{0.070}{0.014} \\
        $K=64$  & \resultnof{0.122}{0.111} & \resultnof{0.243}{0.084} & \resultnof{0.641}{0.120} & \resultnof{0.284}{0.098} & \resultnof{0.059}{0.075} \\
        $K=128$  & \resultnof{0.150}{0.095} & \resultnof{0.258}{0.097} & \resultnof{0.672}{0.101} & \resultnof{0.291}{0.082} & \resultnof{0.050}{0.042} \\
        $K=256$  & \resultnof{0.161}{0.064} & \resultnof{0.263}{0.058} & \resultnof{0.705}{0.068} & \resultnof{0.251}{0.061} & \resultnof{0.030}{0.033} \\
        \rowcolor{gray!10}
        $K=512$  & \resultnof{0.140}{0.045} & \resultnof{0.238}{0.035} & \resultnof{0.728}{0.055} & \resultnof{0.275}{0.074} & \resultnof{0.025}{0.043} \\

        \midrule
        \multicolumn{6}{c}{\textbf{Embedding dimensionality $d$ with $K=32$} }\\
        \midrule
        $d=64$  & \resultnof{0.249}{0.026} & \resultnof{0.365}{0.038} & \resultnof{0.660}{0.184} & \resultnof{0.263}{0.023} & \resultnof{0.020}{0.040} \\

        $d=128$ & \resultnof{0.128}{0.015} & \resultnof{0.231}{0.012} & \resultnof{0.657}{0.018} & \resultnof{0.295}{0.017} & \resultnof{0.070}{0.014} \\
        $d=256$  & \resultnof{0.203}{0.088} & \resultnof{0.325}{0.073} & \resultnof{0.714}{0.112} & \resultnof{0.294}{0.087} & \resultnof{0.024}{0.051} \\
        \rowcolor{gray!10}

        \bottomrule
    \end{tabular}}
    \label{supp:tuning_forecasting}
\end{table*}

\begin{table*}[htb!]
    \centering
        \caption{\textbf{Classification performance for varying values of $K$ and $d$.} }

    \footnotesize {
    \begin{tabular}{c|cc}
        \toprule
        \multicolumn{3}{c}{\textbf{Classification}}\\
        \midrule
        \textbf{Value} & \textbf{Acc} ($\uparrow$)& \textbf{$F_1$} ($\uparrow$)\\
        \midrule
        \multicolumn{3}{c}{\textbf{Quantization indexes $K$ with $d$ = 128}}\\
        \midrule
        $K=4$  & \resultnof{78.03}{4.39} & \resultnof{66.64}{7.27}\\
        \rowcolor{gray!10}
        $K=8$  & \resultnof{78.04}{4.08} & \resultnof{66.90}{7.07}\\
        $K=16$  & \resultnof{78.20}{4.39} & \resultnof{66.96}{7.25}\\
        \rowcolor{gray!10}
        $K=32$  & \resultnof{77.96}{4.33} & \resultnof{66.06}{8.32}\\
        $K=64$  & \resultnof{78.01}{4.35} & \resultnof{66.12}{7.25}\\
        \rowcolor{gray!10}
        $K=128$  & \resultnof{78.18}{4.37} & \resultnof{66.88}{7.18}\\
        $K=256$  & \resultnof{78.01}{4.37} & \resultnof{67.02}{6.90}\\
        \rowcolor{gray!10}
        $K=512$  & \resultnof{77.95}{4.52} & \resultnof{66.78}{7.14}\\
        
        \midrule
        \multicolumn{3}{c}{\textbf{Embedding dimensionality $d$ with $K$ = 32}}\\
        \midrule
        $d=64$  & \resultnof{78.04}{4.13} & \resultnof{66.89}{6.93}\\
        \rowcolor{gray!10}
        $d=128$  & \resultnof{77.96}{4.33} & \resultnof{66.06}{8.32}\\
        $d=256$  & \resultnof{77.79}{4.55} & \resultnof{66.57}{7.08}\\

        \bottomrule
    \end{tabular}}
    \label{supp:tuning_classification}
\end{table*}

The optimal values for $K$ and $d$ were decided by the highest value of Pearson correlation obtained in the downstream task of forecasting (Table~\ref{supp:tuning_forecasting}, best correlation obtained for values $K$ = 32 and $d$ = 128).  Table~\ref{supp:tuning_classification}, instead, shows the performance obtained in the classification downstream task for varying values of $K$ and $d$.

\subsection{Additional classification diagnostics under class imbalance}

\qftext{As an additional diagnostic for the response-classification task under class imbalance, we report AUROC and AUPRC on the drifting gratings stimulus in Table~\ref{tab:response_classification_drifting}. We use this setting as a representative benchmark. Balanced accuracy remains the primary classification metric in the main experiments, since predictions are evaluated at a fixed decision threshold and the metric accounts for both responsive and non-responsive classes. AUROC and AUPRC provide complementary threshold-independent diagnostics, with AUPRC being particularly informative for the minority responsive-neuron class.}

\qftext{The results confirm the same trend observed with balanced accuracy. \ours achieves the best performance across all three metrics, with AUROC of $0.9222 \pm 0.0327$, AUPRC of $0.7032 \pm 0.0882$, and balanced accuracy of $0.7796 \pm 0.0433$. The improvement in AUPRC is especially relevant under class imbalance, indicating that \ours better preserves precision--recall trade-offs when identifying responsive neurons.}

\begin{table*}[htbp]
\centering
\caption{\textbf{Additional response-classification metrics on drifting gratings.} AUROC and AUPRC are reported as complementary threshold-independent metrics under class imbalance.}
\label{tab:response_classification_drifting}
\rowcolors{2}{gray!10}{white}
\begin{tabular}{lccc}
\toprule
\textbf{Model} & \textbf{AUROC} & \textbf{AUPRC} & \textbf{Balanced accuracy} \\
\midrule
LSTM         & $0.5024 \pm 0.0076$ & $0.0449 \pm 0.0145$ & $0.4986 \pm 0.0056$ \\
Informer     & $0.7057 \pm 0.0714$ & $0.2734 \pm 0.1448$ & $0.5852 \pm 0.0603$ \\
Autoformer   & $0.6307 \pm 0.0560$ & $0.1633 \pm 0.1166$ & $0.5704 \pm 0.0564$ \\
Crossformer  & $0.7983 \pm 0.0751$ & $0.3570 \pm 0.1709$ & $0.5819 \pm 0.0689$ \\
BrainLM & $0.7934 \pm 0.1460$ & $0.5497 \pm 0.2595$ & $0.7177 \pm 0.1167$ \\
BrainLM$_{ft}$   & $0.6275 \pm 0.0217$ & $0.0763 \pm 0.0211$ & $0.4991 \pm 0.0048$ \\
\textbf{\ours} & $\mathbf{0.9222 \pm 0.0327}$ & $\mathbf{0.7032 \pm 0.0882}$ & $\mathbf{0.7796 \pm 0.0433}$ \\
\bottomrule
\end{tabular}
\end{table*}

\subsection{Effect-size analysis for forecasting metrics}

\qftext{To complement the statistical significance tests reported in the main paper, we additionally report Cohen's $d$ effect sizes for the forecasting metrics in Table~\ref{tab:cohensd}. This analysis addresses whether the observed improvements correspond to practically meaningful standardized differences, especially in cases where the absolute metric difference may appear small relative to the cross-subject variability.}

\qftext{Cohen's $d$ is computed between \ours and each competing method for each forecasting metric. For interpretability, the sign is oriented so that positive values always indicate an advantage of \ours: for MSE and SMAPE, where lower is better, the sign is inverted, while for correlation and SSIM, where higher is better, the standard direction is retained. Following common conventions, values around $0.2$, $0.5$, and $0.8$ indicate small, medium, and large effects, respectively.}

\qftext{The effect-size analysis shows that the absolute MSE improvement over Informer corresponds to a small standardized effect ($d=0.068$), consistent with the large variability of squared-error values across subjects and stimuli. However, the same comparison shows substantially larger positive effects for metrics that better capture temporal structure and relative forecasting quality: $d=0.574$ for correlation and $d=1.086$ for SMAPE, where positive values indicate an advantage of \ours. This indicates that, although the MSE difference is numerically small, the gains of \ours are more evident in metrics that reflect temporal alignment and normalized prediction error. Overall, these results support the interpretation that the learned quantization pipeline improves the forecasting of neural dynamics beyond absolute scale differences.}

\begin{table*}[htb!]
    \centering
    \renewcommand{\arraystretch}{1.2}
        \caption{\qftext{\textbf{Cohen's $d$ effect sizes for forecasting metrics.}
        Effect sizes are computed between \ours and each competing method, with signs oriented so that positive values indicate better performance of \ours.}}
        \label{tab:cohensd}
    \begin{tabular}{lrrrr}
    \toprule
         Method &   MSE &  SMAPE &  Corr. &  SSIM \\
    \midrule
           LSTM & 0.333 & 10.252 & -0.225 & 3.042 \\
     Autoformer & 0.704 &  2.036 &  0.290 & 2.181 \\
       Informer & 0.068 &  1.086 &  0.574 & 1.682 \\
        BrainLM & 0.282 &  0.353 &  0.008 & 1.249 \\
     BrainLM$_{ft}$ & 0.245 &  0.539 & -0.160 & 1.509 \\
    Crossformer & 0.688 &  1.022 &  0.526 & 0.917 \\
    \bottomrule
    \end{tabular}
\end{table*}

\qftext{
\subsection{Forecasting architectures, training budgets, and additional baselines}
\label{sec:forecasting_baselines}

\noindent\textbf{Additional causal convolutional baselines.} To complement the transformer-based and recurrent competitors, we additionally evaluate two causal convolutional baselines: a Temporal Convolutional Network (TCN)~\cite{bai2018tcn} and a one-dimensional Seq-U-Net~\cite{stoller2020sequnet}. Both models receive the same pre-stimulus neural activity and stimulus descriptor used by the other forecasting methods. Neurons are represented as explicit input channels, while the stimulus descriptor is projected into the latent channel space and added to the temporal representation. A shared temporal head maps the 96 observed frames to the 64-frame forecast.

The TCN uses five causal temporal-convolution levels with 128 hidden channels, kernel size 3, and dropout 0.2. Its receptive field covers all 96 observed frames. Seq-U-Net uses five causal encoder--decoder levels with skip connections, the same hidden width, kernel size, and dropout as the TCN, providing a controlled comparison between two different causal convolutional inductive biases.

Both baselines are trained with Adam for 100 downstream epochs using a batch size of 32 and a learning rate of $10^{-4}$. Training is performed independently for each container--stimulus pair using the same chronological splits and the same observed neural activity and stimulus information adopted for the other methods. Table~\ref{tab:additional_convolutional_baselines} compares TCN and Seq-U-Net with \ours under the original accumulated-gradient evaluation protocol described in the \hyperref[supp:forecasting_norm_peak]{following Section}.

\begin{table*}[h]
    \centering
    \caption{\textbf{Forecasting performance of the additional convolutional baselines under the accumulated-gradient evaluation described in the \hyperref[supp:forecasting_norm_peak]{following Section}.} TCN, Seq-U-Net, and QuantFormer were trained for 100 downstream epochs. Values are reported as mean $\pm$ standard deviation across container--stimulus pairs. Peak metrics are evaluated on active observations only. Peak timing is reported in seconds, and peak-amplitude MAE is computed after accumulated-gradient normalization. Best mean results are highlighted in bold.}
    \footnotesize
    \resizebox{\linewidth}{!}{%
    \begin{tabular}{l|cccccc}
        \toprule
        \textbf{Method} & \textbf{MSE} ($\downarrow$) & \textbf{SMAPE} ($\downarrow$) & \textbf{Corr} ($\uparrow$) & \textbf{SSIM} ($\uparrow$) & \textbf{Peak timing} ($\downarrow$) & \textbf{Peak amplitude} ($\downarrow$) \\
        \midrule
        \rowcolor{gray!10}
        TCN & $0.480 \pm 0.371$ & $0.733 \pm 0.017$ & $0.337 \pm 0.067$ & $0.030 \pm 0.029$ & $0.320 \pm 0.215$ & $0.616 \pm 0.276$ \\
        Seq-U-Net & $0.392 \pm 0.290$ & $0.744 \pm 0.015$ & $0.318 \pm 0.063$ & $0.027 \pm 0.019$ & $0.341 \pm 0.212$ & $0.578 \pm 0.253$ \\
        \midrule
        \rowcolor{gray!10}
        \textbf{\ours} & $\mathbf{0.247 \pm 0.078}$ & $\mathbf{0.656 \pm 0.137}$ & $\mathbf{0.338 \pm 0.075}$ & $\mathbf{0.069 \pm 0.062}$ & $\mathbf{0.272 \pm 0.179}$ & $\mathbf{0.417 \pm 0.147}$ \\
        \bottomrule
    \end{tabular}}
    \label{tab:additional_convolutional_baselines}
\end{table*}

\ours obtains the lowest mean MSE and SMAPE, the highest mean correlation, and the lowest peak-timing and peak-amplitude errors among the three methods. TCN and Seq-U-Net nevertheless provide competitive causal convolutional references, indicating that the performance of \ours under this protocol is not explained solely by the use of causal temporal modeling.

\noindent\textbf{Training budgets and architectural adaptations.} All downstream methods are trained for 100 epochs using the same chronological train/test splits and no validation split. QuantFormer's 50-epoch self-supervised pretraining and the external pretraining used by BrainLM$_{ft}$ are reported separately from downstream optimization. 
Table~\ref{tab:forecasting_budget_adaptations} reports model architectures, trainable parameter ranges, optimization settings, initialization or pretraining strategies, and neural-data adaptations. Parameter ranges reflect differences in neuron count and stimulus-descriptor dimensionality across container--stimulus models.

\begin{table*}[htbp]
\centering
\caption{\textbf{Forecasting training budget and architectural adaptations.} Trainable parameter counts are reported as the minimum--maximum range across container--stimulus models; the variation reflects the container-specific neuron count and stimulus-descriptor dimensionality. All trainable methods use the same chronological train/test split and no validation split. All methods use 100 downstream epochs. Architecture-specific pretraining is reported separately from downstream optimization.}
\label{tab:forecasting_budget_adaptations}
\scriptsize
\setlength{\tabcolsep}{2pt}
\renewcommand{\arraystretch}{0.92}
\begin{tabularx}{\textwidth}{
l
>{\hsize=1.00\hsize\linewidth=\hsize}X
c
>{\hsize=0.75\hsize\linewidth=\hsize}X
>{\hsize=0.75\hsize\linewidth=\hsize}X
>{\hsize=1.50\hsize\linewidth=\hsize}X
}
\toprule
\textbf{Method} & \textbf{Architecture} & \textbf{Trainable params} & \makecell[l]{\textbf{Downstream}\\\textbf{optimization}} &
\makecell[l]{\textbf{Initialization /}\\\textbf{pretraining}} & \textbf{Task adaptation} \\
\midrule
Mean baseline & Observed-baseline mean repeated over the forecast horizon & 0 & -- & None & Uses only the observed pre-stimulus interval; no learned parameters. \\
LSTM & 2 LSTM layers; hidden size 32 & 41,331--871,399 & Adam; 100 epochs; batch 32; lr=0.0001; wd=0.0005; MSE & Random initialization & Container-specific multivariate input/output sizes; stimulus descriptor concatenated before the output projection; autoregressive 64-frame decoding. \\
TCN & 5 causal temporal-convolution levels; 128 channels; kernel 3; dropout 0.2 & 538,835--854,951 & Adam; 100 epochs; batch 32; lr=0.0001; wd=0.0005; MSE & Random initialization & Neurons are explicit container-specific channels; 1x1 input/output channel projections; projected stimulus added to the latent sequence; shared 96-to-64 temporal head. \\
Seq-U-Net & 5 causal encoder-decoder levels; 128 channels; kernel 3; dropout 0.2 & 1,095,635--1,411,751 & Adam; 100 epochs; batch 32; lr=0.0001; wd=0.0005; MSE & Random initialization & Neurons are explicit container-specific channels; 1x1 input/output channel projections; projected stimulus added to the latent sequence; shared 96-to-64 temporal head. \\
Autoformer & 2 encoder + 2 decoder layers; d=128; 4 heads; d\_ff=512 & 1,204,262--1,873,358 & Adam; 100 epochs; batch 128; lr=0.0001; wd=0.0005; MSE & Random initialization & Container-specific input/output projections; stimulus descriptor projected to d and fused before the final forecast projection; 64-frame direct output. \\
Informer & 2 encoder + 2 decoder layers; d=128; 4 heads; d\_ff=512 & 1,146,790--1,653,070 & Adam; 100 epochs; batch 128; lr=0.0001; wd=0.0005; MSE & Random initialization & Container-specific input/output projections; stimulus descriptor projected to d and fused before the final forecast projection; 64-frame direct output. \\
BrainLM & 4 encoder + 2 decoder layers; d=512; 4 heads; d\_ff=1024 & 13,966,912--15,013,440 & Adam; 100 epochs; batch 32; lr=0.001; wd=0.0005; MSE & Random initialization & fMRI voxel dimension replaced by the container's neuron count; mouse neuron coordinates supplied through the XYZ pathway; stimulus-conditioned decoder; 60-frame output required by time-patch size 20. \\
BrainLM$_{ft}$ & 4 encoder + 2 decoder layers; d=512; 4 heads; d\_ff=1024 & 13,966,912--15,013,440 & Adam; 100 epochs; batch 32; lr=0.001; wd=0.0005; MSE & External BrainLM checkpoint; end-to-end fine-tuning & Same neural-data adapter as BrainLM; all pretrained encoder and decoder parameters fine-tuned end to end. \\
Crossformer & 2 encoder scales; d=128; 8 heads; d\_ff=512; segment length 16 & 2,901,552--3,434,544 & Adam; 100 epochs; batch 32; lr=0.0001; wd=0.0005; MSE & Random initialization & Neurons represented as container-specific dimensions; learned stimulus projection added to decoder queries; 96 observed frames mapped directly to 64 forecast frames. \\
QuantFormer & 6 encoder layers; d=128; 8 heads; d\_ff=512; patch 16; K=32 & 1,217,056--1,505,824 & Adam; 100 epochs; batch 32; lr=0.001; wd=0.0005; CE & 50-epoch self-supervised pretraining; downstream fine-tuning of the encoder, neuron/stimulus prompts, and code-index classification head; codebook and reconstruction decoder frozen & Learned neuron and stimulus prompts are prepended to the patch tokens; the four masked response patches are classified into codebook indices and reconstructed using the frozen codebook and decoder. \\
\bottomrule
\end{tabularx}
\end{table*}
}

\qftext{
\subsection{Forecasting evaluation across normalization protocols and peak-based metrics}
\label{supp:forecasting_norm_peak}

We evaluate forecasting performance under three complementary normalization protocols: (i) the accumulated-gradient normalization used for the main forecasting results, (ii) evaluation on the baseline-normalized $\Delta F/F$ responses without additional evaluation-time normalization, and (iii) two prediction-independent per-neuron normalizations based exclusively on training-set statistics. We additionally introduce peak-timing and peak-amplitude metrics to directly evaluate stimulus-evoked fluorescence transients. These evaluations characterize different aspects of forecasting performance and should therefore be interpreted separately.

\noindent\textbf{Accumulated-gradient normalization.} Calcium fluorescence responses are temporally sparse, with extended near-baseline intervals and brief stimulus-evoked transients. Because conventional point-wise regression metrics assign equal weight to every temporal position, their aggregate values can be dominated by the non-transient portions of the response window, allowing conservative predictions close to the baseline to obtain competitive MSE or MAE despite missing stimulus-evoked transients. We therefore use accumulated-gradient normalization to reduce sensitivity to absolute transient amplitude and evaluate agreement in temporal variation.

For a response trace $\mathbf{x}_f$ and its forecast $\widehat{\mathbf{x}}_f$, we define, for $\mathbf{v}\in\{\mathbf{x}_f,\widehat{\mathbf{x}}_f\}$,
\begin{equation}
G(\mathbf{v})=\sum_{t=1}^{L_f}\left|\nabla_t\mathbf{v}\right|,\qquad \mathcal{N}(\mathbf{v})=\frac{\mathbf{v}}{G(\mathbf{v})},
\end{equation}
where $\nabla_t\mathbf{v}$ denotes the discrete temporal derivative at time $t$. Target and prediction are normalized independently:
\begin{equation}
\widetilde{\mathbf{x}}_f=\mathcal{N}(\mathbf{x}_f),\qquad \widetilde{\widehat{\mathbf{x}}}_f=\mathcal{N}(\widehat{\mathbf{x}}_f).
\end{equation}

By normalizing targets and predictions independently, this protocol reduces sensitivity to absolute transient amplitude and evaluates agreement in normalized temporal dynamics rather than absolute forecasting accuracy on the baseline-normalized $\Delta F/F$ scale. Accordingly, the main forecasting table reports performance under this specific evaluation protocol and should not be interpreted as an unconditional ranking of point-wise reconstruction accuracy.

\noindent\textbf{Evaluation without additional normalization.} Table~\ref{tab:forecasting_un-normalized} reports forecasting metrics computed directly on the baseline-normalized $\Delta F/F$ responses, without the additional accumulated-gradient normalization. These results provide a measure of point-wise reconstruction accuracy in the fluorescence-response scale used by the models.

The mean-signal baseline obtains competitive point-wise errors because non-transient temporal positions occupy a large portion of the response window. Errors at these positions therefore make a larger aggregate contribution to MSE and MAE than errors occurring during brief stimulus-evoked transients.

\begin{table*}[htb!]
    \centering
        \caption{\textbf{Forecasting performance without additional evaluation-time normalization.} Metrics are computed directly on baseline-normalized $\Delta F/F$ responses. Values are reported as mean $\pm$ standard deviation across container--stimulus pairs.}
    \footnotesize {
    \begin{tabular}{ll|ccccc}
        \toprule
        &\textbf{Method} & \textbf{MSE} ($\downarrow$) &    
        \textbf{MAE} ($\downarrow$) & \textbf{SMAPE} ($\downarrow$) & \textbf{Corr} ($\uparrow$) & \textbf{SSIM} ($\uparrow$) \\
        \midrule
        \rowcolor{gray!10}
        & Baseline & \resultnof{0.095}{0.341} & \resultnof{0.058}{0.008} & \resultnof{0.829}{0.009} & \resultnof{0.335}{0.002} & \resultnof{0.122}{0.031}\\ 
        & LSTM  & \resultnof{0.093}{0.395} & \resultnof{0.505}{0.007} & \resultnof{0.883}{0.062} & \resultnof{0.252}{0.322} & \resultnof{0.246}{0.026}\\
            \rowcolor{gray!10}
        & Autoformer & \resultnof{0.098}{0.123} & \resultnof{0.074}{0.022} & \resultnof{0.062}{0.025} & \resultnof{0.118}{0.011} & \resultnof{0.077}{0.037}\\ 
        & Informer & \resultnof{0.097}{0.379} & \resultnof{0.062}{0.010} & \resultnof{0.857}{0.049} & \resultnof{0.118}{0.011} & \resultnof{0.098}{0.032}\\ 
    \rowcolor{gray!10}

        & BrainLM & \resultnof{0.103}{0.388} & \resultnof{0.057}{0.008} & \resultnof{0.902}{0.049} & \resultnof{0.106}{0.006} & \resultnof{0.111}{0.031}\\  
        & BrainLM$_{ft}$ & \resultnof{0.132}{0.451} & \resultnof{0.057}{0.008} & \resultnof{0.858}{0.057} & \resultnof{0.107}{0.073} & \resultnof{0.098}{0.042} \\
        \rowcolor{gray!10}
        & Crossformer & \resultnof{0.301}{1.210} & \resultnof{0.060}{0.009} & \resultnof{0.771}{0.039} & \resultnof{0.138}{0.032}& \resultnof{0.096}{0.032}\\ 
        \midrule
        \rowcolor{gray!10}
        & \textbf{\ours} & \resultnof{0.445}{1.230} & \resultnof{0.236}{0.106} & \resultnof{1.55}{0.082} & \resultnof{0.138}{0.017} & \resultnof{0.015}{0.022}\\
        \bottomrule
    \end{tabular}}
    \label{tab:forecasting_un-normalized}
\end{table*}

\noindent\textbf{Prediction-independent normalization.} To complement the accumulated-gradient evaluation, we repeated the forecasting analysis using two prediction-independent per-neuron normalization protocols. For each container--stimulus pair, the normalization statistics were estimated using all valid response samples from the training trials. The same training-derived statistics were applied to targets and predictions. No normalization statistics were estimated from test responses or model outputs.

For the per-neuron $z$-score protocol, the target response and corresponding prediction are normalized as
\begin{equation}
z^{(i)}_{f,n,t}=\frac{x^{(i)}_{f,n,t}-\mu_n^{\mathrm{train}}}{\max\!\left(\sigma_n^{\mathrm{train}},\epsilon\right)},\qquad \widehat{z}^{(i)}_{f,n,t}=\frac{\widehat{x}^{(i)}_{f,n,t}-\mu_n^{\mathrm{train}}}{\max\!\left(\sigma_n^{\mathrm{train}},\epsilon\right)}.
\end{equation}
Here, $i$, $n$, and $t$ index the trial, neuron, and valid temporal position, respectively. The quantities $\mu_n^{\mathrm{train}}$ and $\sigma_n^{\mathrm{train}}$ denote the training-set mean and standard deviation of neuron $n$, while $\epsilon$ is a small constant introduced for numerical stability.

We also consider a robust per-neuron normalization based on the training-set median and interquartile range:
\begin{equation}
r^{(i)}_{f,n,t}=\frac{x^{(i)}_{f,n,t}-m_n^{\mathrm{train}}}{\max\!\left(\mathrm{IQR}_n^{\mathrm{train}},\epsilon\right)},\qquad \widehat{r}^{(i)}_{f,n,t}=\frac{\widehat{x}^{(i)}_{f,n,t}-m_n^{\mathrm{train}}}{\max\!\left(\mathrm{IQR}_n^{\mathrm{train}},\epsilon\right)}.
\end{equation}
where $m_n^{\mathrm{train}}$ and $\mathrm{IQR}_n^{\mathrm{train}}$ are, respectively, the median and interquartile range computed exclusively from the training responses of neuron $n$.

Unlike accumulated-gradient normalization, these two transformations do not depend on the model output. The same training-derived affine transformation is applied to each target and its corresponding prediction. Results are reported in Tables~\ref{tab:forecast_train_norm} and~\ref{tab:forecast_train_robust}.

\noindent\textbf{Peak-based transient metrics.} To complement the point-wise forecasting metrics, we introduce two metrics that directly characterize stimulus-evoked fluorescence transients. For each active observation, the peak-timing error is defined as
\begin{equation}
E_{\mathrm{time}}=\frac{1}{f_s}\left|\arg\max_t x_{f,t}-\arg\max_t \widehat{x}_{f,t}\right|,
\end{equation}
where $f_s$ is the sampling frequency. Peak-timing error is reported in seconds.

The peak-amplitude error is defined as
\begin{equation}
E_{\mathrm{amp}}=\left|\max_t x_{f,t}-\max_t \widehat{x}_{f,t}\right|.
\end{equation}
Peak-amplitude MAE is computed using the same prediction-independent per-neuron normalization adopted for the corresponding evaluation table. It is therefore expressed either in the per-neuron $z$-score scale or in the per-neuron median/IQR scale.

Both peak metrics are evaluated on active observations only. They operate directly on the fluorescence responses and do not require spike deconvolution.

\begin{table*}[htbp]
\centering
\renewcommand{\arraystretch}{1.2}
\caption{\textbf{Forecasting performance under prediction-independent per-neuron $z$-score normalization.} Per-neuron means and standard deviations used for normalization are estimated exclusively from training responses and applied to both targets and predictions. Values are reported as mean $\pm$ standard deviation across container--stimulus pairs.
Peak metrics are evaluated on active observations only. Best mean results are highlighted in bold.}
\label{tab:forecast_train_norm}
\resizebox{\textwidth}{!}{%
\begin{tabular}{l|cccccc}
\toprule
\textbf{Method} & \textbf{MSE} $(\downarrow)$ & \textbf{SMAPE} $(\downarrow)$ & \textbf{Corr.} $(\uparrow)$ & \textbf{SSIM} $(\uparrow)$ & \textbf{Peak timing (s)} $(\downarrow)$ & \textbf{Peak amplitude MAE} $(\downarrow)$ \\
\midrule
\rowcolor{gray!10}
Mean-signal baseline & 4.328 $\pm$ 1.224 & 0.934 $\pm$ 0.030 & 0.000 $\pm$ 0.000 & 0.002 $\pm$ 0.003 & 0.340 $\pm$ 0.253 & 3.288 $\pm$ 0.735 \\
LSTM          & 4.292 $\pm$ 1.219 & 0.904 $\pm$ 0.063 & 0.179 $\pm$ 0.122 & 0.001 $\pm$ 0.003 & 0.333 $\pm$ 0.244 & 3.301 $\pm$ 0.759 \\
\rowcolor{gray!10}
TCN & 4.548 $\pm$ 1.947 & 0.877 $\pm$ 0.031 & 0.269 $\pm$ 0.047 & 0.004 $\pm$ 0.005 & 0.311 $\pm$ 0.216 & 3.277 $\pm$ 0.729 \\
Seq-U-Net & 4.643 $\pm$ 1.801 & 0.912 $\pm$ 0.032 & 0.255 $\pm$ 0.047 & 0.001 $\pm$ 0.003 & 0.328 $\pm$ 0.214 & 3.291 $\pm$ 0.739 \\
\rowcolor{gray!10}
Autoformer    & 4.920 $\pm$ 1.472 & 0.774 $\pm$ 0.026 & 0.253 $\pm$ 0.072 & $0.000$ $\pm$ 0.002 & 0.330 $\pm$ 0.229 & 3.165 $\pm$ 0.597 \\
Informer      & 4.338 $\pm$ 1.226 & 0.885 $\pm$ 0.039 & 0.238 $\pm$ 0.055 & 0.001 $\pm$ 0.002 & 0.302 $\pm$ 0.221 & 3.220 $\pm$ 0.681 \\
BrainLM       & 4.316 $\pm$ 1.225 & 0.918 $\pm$ 0.036 & \textbf{0.271 $\pm$ 0.059} & 0.002 $\pm$ 0.003 & 0.329 $\pm$ 0.237 & 3.314 $\pm$ 0.745 \\
\rowcolor{gray!10}
BrainLM$_{\mathrm{ft}}$ & 3.955 $\pm$ 0.781 & 0.911 $\pm$ 0.034 & 0.259 $\pm$ 0.069 & 0.002 $\pm$ 0.004 & 0.371 $\pm$ 0.262 & 3.304 $\pm$ 0.775 \\
Crossformer   & \textbf{3.757 $\pm$ 1.146} & 0.704 $\pm$ 0.077 & 0.245 $\pm$ 0.053 & 0.002 $\pm$ 0.005 & 0.285 $\pm$ 0.209 & 3.005 $\pm$ 0.744 \\
\midrule
\rowcolor{gray!10}
\textbf{\ours} & 5.599 $\pm$ 2.371 & \textbf{0.671 $\pm$ 0.048} & 0.256 $\pm$ 0.066 & \textbf{0.024 $\pm$ 0.016} & \textbf{0.263 $\pm$ 0.177} & \textbf{2.494 $\pm$ 0.788} \\
\bottomrule
\end{tabular}}
\end{table*}

\begin{table*}[htbp]
\centering
\renewcommand{\arraystretch}{1.2}
\caption{\textbf{Forecasting performance under prediction-independent per-neuron median/IQR normalization.} Per-neuron medians and interquartile ranges are estimated exclusively from training responses and applied to both targets and predictions. Values are reported as mean $\pm$ standard deviation across container--stimulus pairs. Peak metrics are evaluated on active observations only. Best mean results are highlighted in bold.}
\label{tab:forecast_train_robust}
\resizebox{\textwidth}{!}{%
\begin{tabular}{l|cccccc}
\toprule
\textbf{Method} & \textbf{MSE} $(\downarrow)$ & \textbf{SMAPE} $(\downarrow)$ & \textbf{Corr.} $(\uparrow)$ & \textbf{SSIM} $(\uparrow)$ & \textbf{Peak timing (s)} $(\downarrow)$ & \textbf{Peak amplitude MAE} $(\downarrow)$ \\
\midrule
\rowcolor{gray!10}
Mean-signal baseline & 10.695 $\pm$ 12.455 & 0.909 $\pm$ 0.019 & 0.000 $\pm$ 0.000 & $0.000$ $\pm$ 0.000 & 0.340 $\pm$ 0.253 & 4.290 $\pm$ 1.055 \\
LSTM & 10.485 $\pm$ 12.361 & 0.869 $\pm$ 0.033 & 0.171 $\pm$ 0.130 & 0.000 $\pm$ 0.001 & 0.337 $\pm$ 0.242 & 4.281 $\pm$ 1.054 \\
\rowcolor{gray!10}
TCN & 25.901 $\pm$ 99.307 & 0.865 $\pm$ 0.021 & 0.269 $\pm$ 0.047 & 0.004 $\pm$ 0.004 & 0.311 $\pm$ 0.216 & 4.251 $\pm$ 0.996 \\
Seq-U-Net & 14.871 $\pm$ 34.851 & 0.882 $\pm$ 0.020 & 0.255 $\pm$ 0.047 & 0.001 $\pm$ 0.002 & 0.328 $\pm$ 0.214 & 4.276 $\pm$ 1.036 \\
\rowcolor{gray!10}
Autoformer & 16.381 $\pm$ 36.604 & 0.772 $\pm$ 0.025 & 0.253 $\pm$ 0.072 & $0.000$ $\pm$ 0.001 & 0.330 $\pm$ 0.229 & 4.188 $\pm$ 0.864 \\
Informer & 10.681 $\pm$ 12.513 & 0.865 $\pm$ 0.030 & 0.238 $\pm$ 0.055 & 0.001 $\pm$ 0.002 & 0.302 $\pm$ 0.221 & 4.217 $\pm$ 0.989 \\
BrainLM & 10.599 $\pm$ 12.482 & 0.883 $\pm$ 0.030 & \textbf{0.271 $\pm$ 0.059} & $0.000$ $\pm$ 0.001 & 0.329 $\pm$ 0.237 & 4.293 $\pm$ 1.051 \\
\rowcolor{gray!10}
BrainLM$_{\mathrm{ft}}$ & 10.923 $\pm$ 14.203 & 0.883 $\pm$ 0.030 & 0.259 $\pm$ 0.069 & 0.000 $\pm$ 0.001 & 0.371 $\pm$ 0.262 & 4.307 $\pm$ 1.116 \\
Crossformer & \textbf{9.341 $\pm$ 12.529} & 0.702 $\pm$ 0.068 & 0.245 $\pm$ 0.053 & 0.000 $\pm$ 0.005 & 0.285 $\pm$ 0.209 & 3.866 $\pm$ 1.017 \\
\midrule
\rowcolor{gray!10}
\textbf{\ours} & 11.184 $\pm$ 12.965 & \textbf{0.673 $\pm$ 0.050} & 0.256 $\pm$ 0.066 & \textbf{0.024 $\pm$ 0.016} & \textbf{0.263 $\pm$ 0.177} & \textbf{3.336 $\pm$ 1.137} \\
\bottomrule
\end{tabular}}
\end{table*}

The $z$-score and median/IQR evaluations, reported in Tables~\ref{tab:forecast_train_norm} and~\ref{tab:forecast_train_robust}, yield the same qualitative pattern. Crossformer achieves the lowest mean MSE, whereas BrainLM achieves the highest mean correlation. \ours obtains the lowest mean SMAPE, peak-timing error, and peak-amplitude error under both prediction-independent normalization protocols.

In particular, \ours achieves a peak-timing error of $0.263$~s under both protocols. Its peak-amplitude MAE is $2.494$ under per-neuron $z$-score normalization and $3.336$ under per-neuron median/IQR normalization. Taken together, these complementary evaluations distinguish between point-wise reconstruction accuracy, agreement in accumulated-gradient-normalized temporal dynamics, and direct evaluation of stimulus-evoked transient timing and amplitude.

}

\pagebreak

\subsection{Response forecasting examples}
\label{suppl:quality}
\qftext{We provide additional examples of response forecasts produced by all tested models across the four stimulus categories, complementing those reported in the main paper} (\emph{Natural scenes} in Figure~\ref{supp_fig:natural}, \emph{Drifting gratings} in Figure~\ref{supp_fig:drifting}, \emph{Static gratings} in Figure~\ref{supp_fig:static} and \emph{Locally sparse noise} in Figure~\ref{supp_fig:noise}). \qftext{These examples qualitatively illustrate the ability of \ours to capture sparse stimulus-evoked fluorescence-response patterns across the four stimulus categories.}

\begin{figure*}[h]
    \centering

    \includegraphics[width=0.45\textwidth]{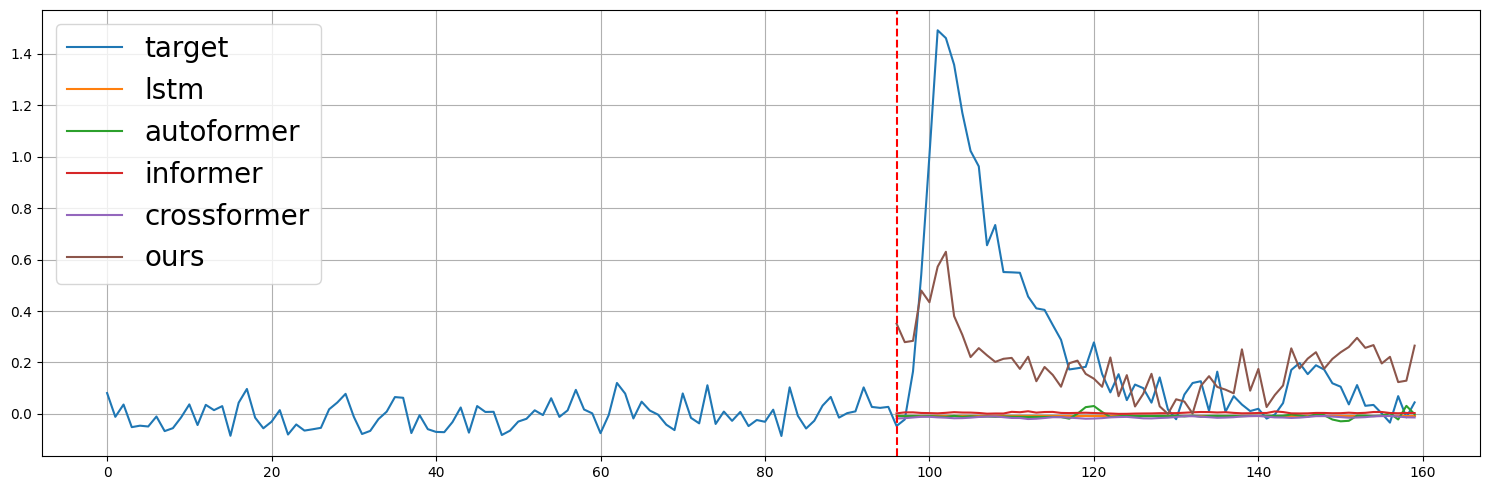}
    \includegraphics[width=0.45\textwidth]{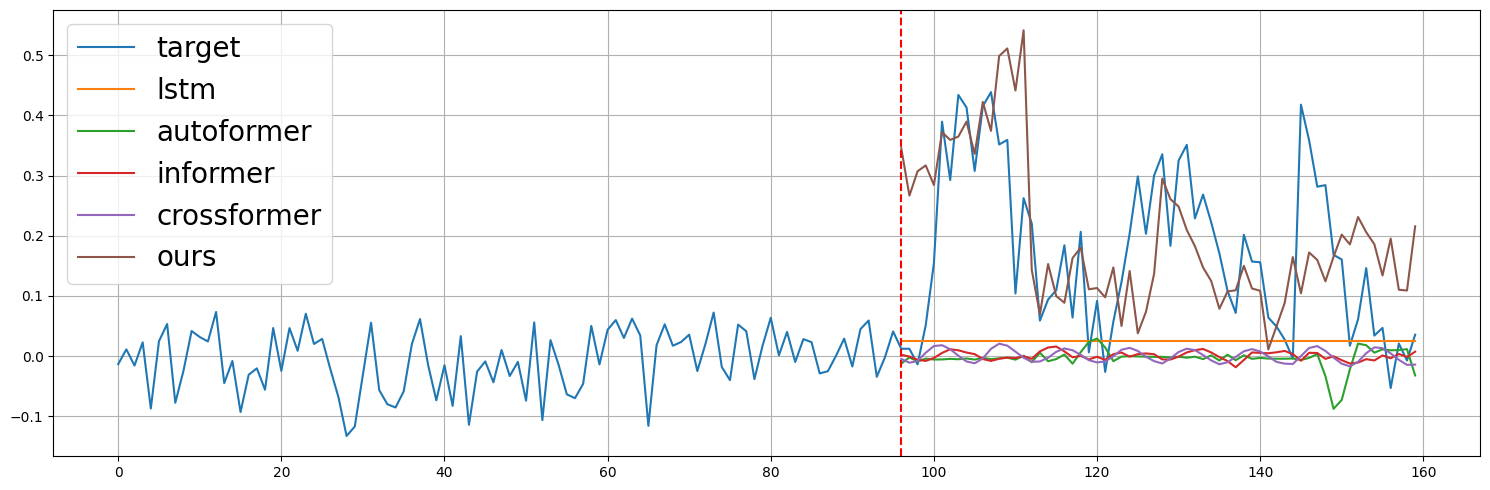}
    \includegraphics[width=0.45\textwidth]{suppl_img/prediction_536323956_natural_scenes_41.png}
    \includegraphics[width=0.45\textwidth]{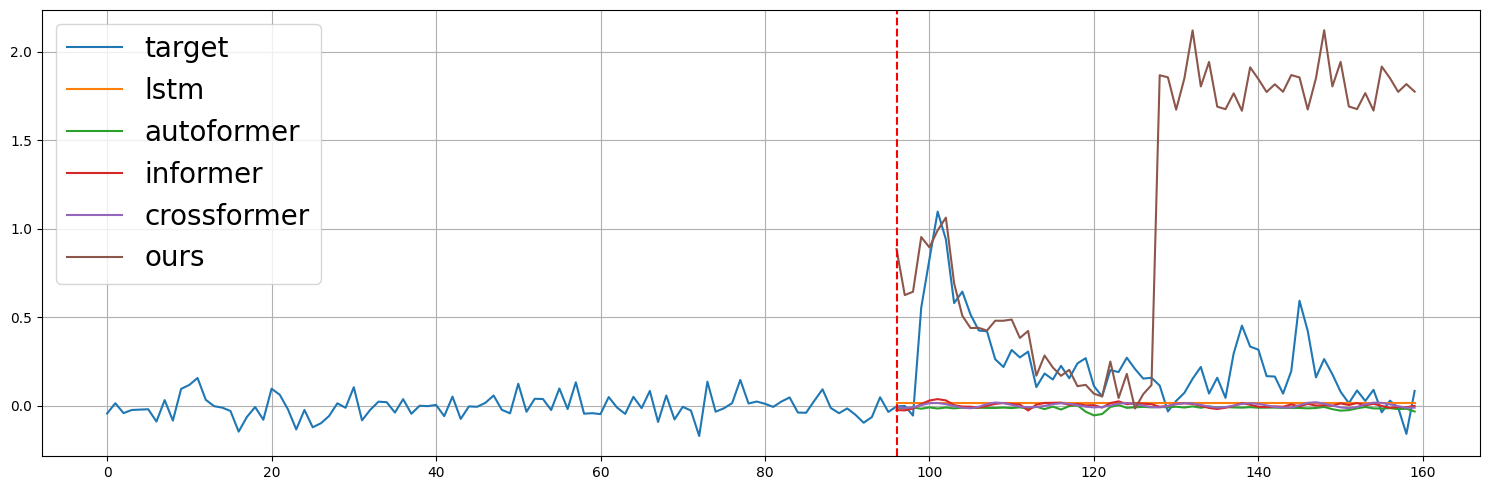}
    \includegraphics[width=0.45\textwidth]{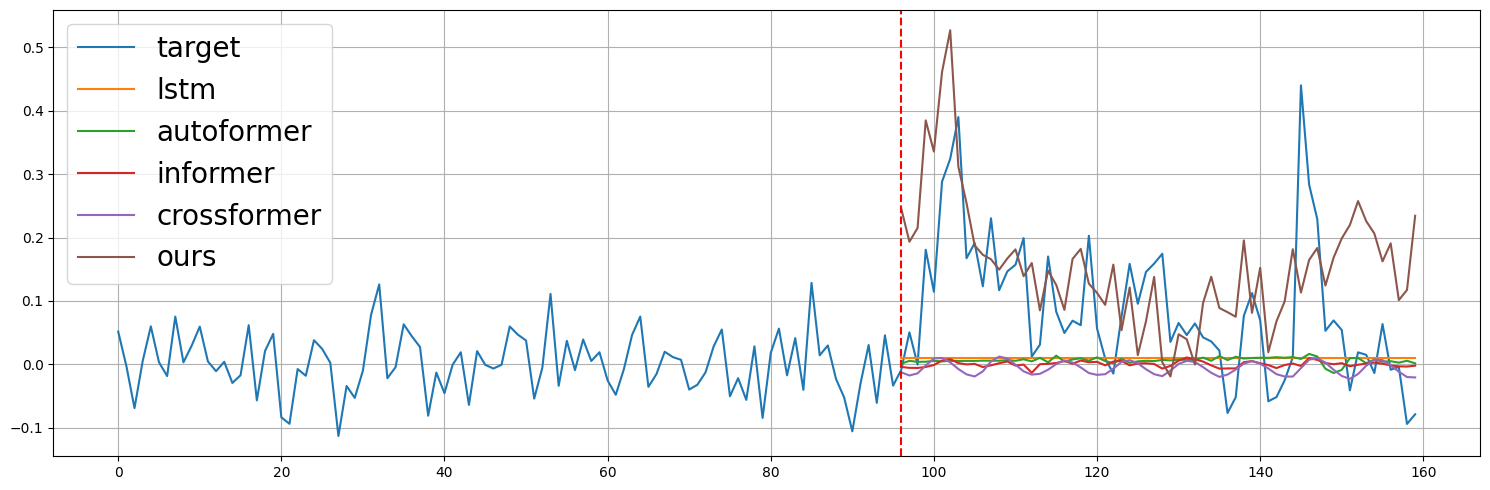}
    \includegraphics[width=0.45\textwidth]{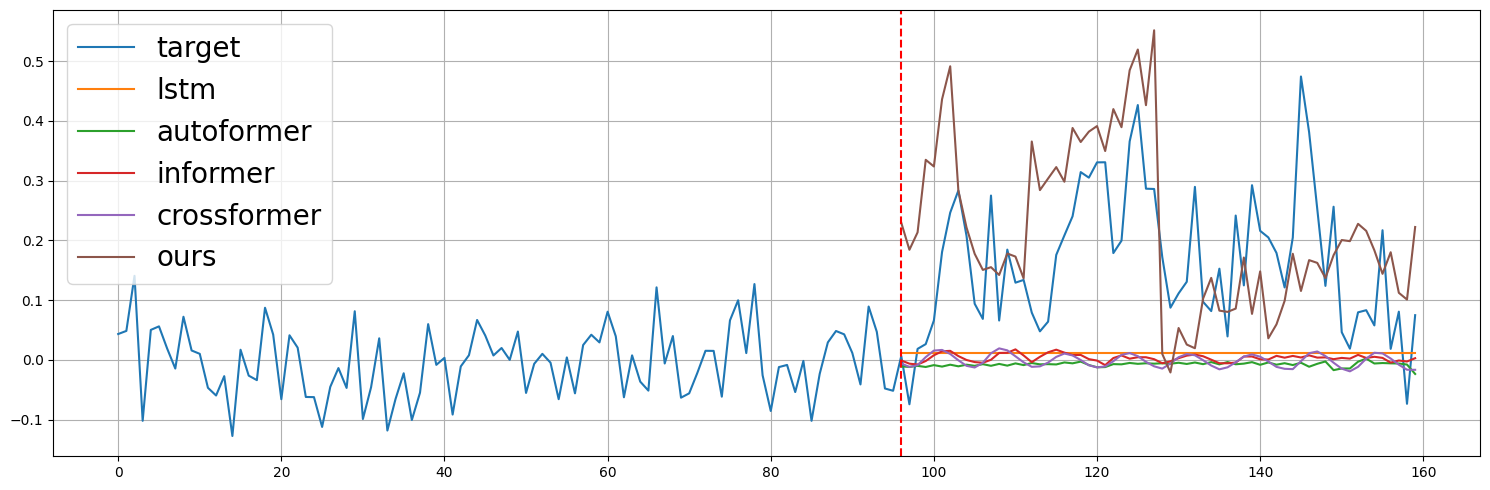}
    \caption{\textbf{Examples of response forecasting by \ours and its competitors on natural scenes}.}
    \label{supp_fig:natural}
\end{figure*}

\begin{figure*}[h]
    \centering
    \includegraphics[width=0.45\textwidth]{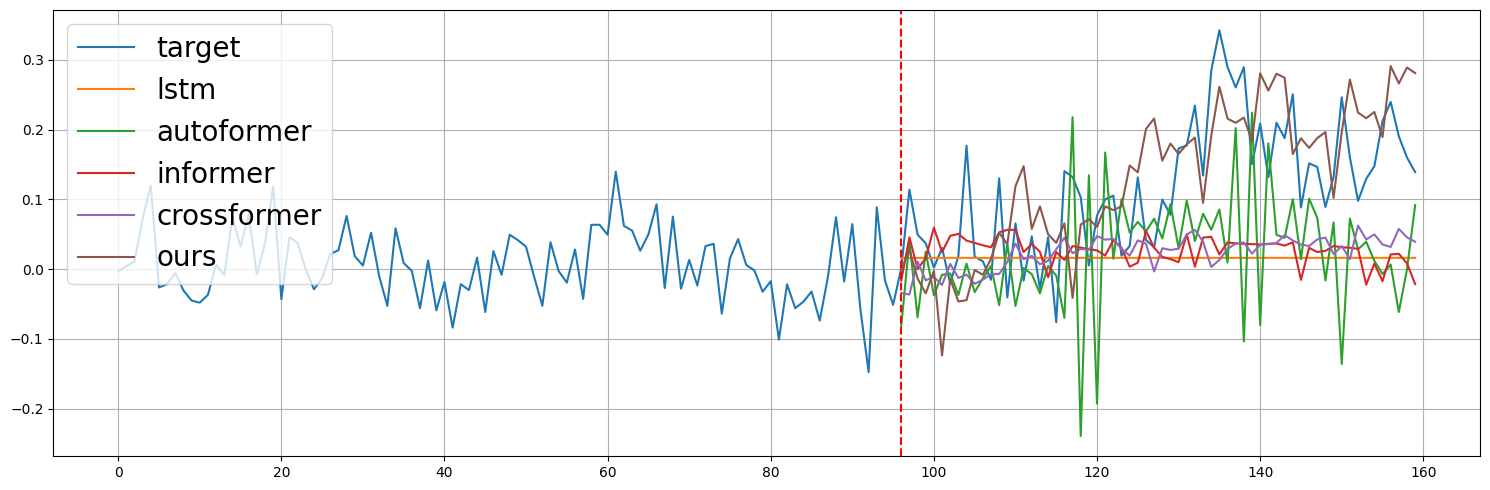}
    \includegraphics[width=0.45\textwidth]{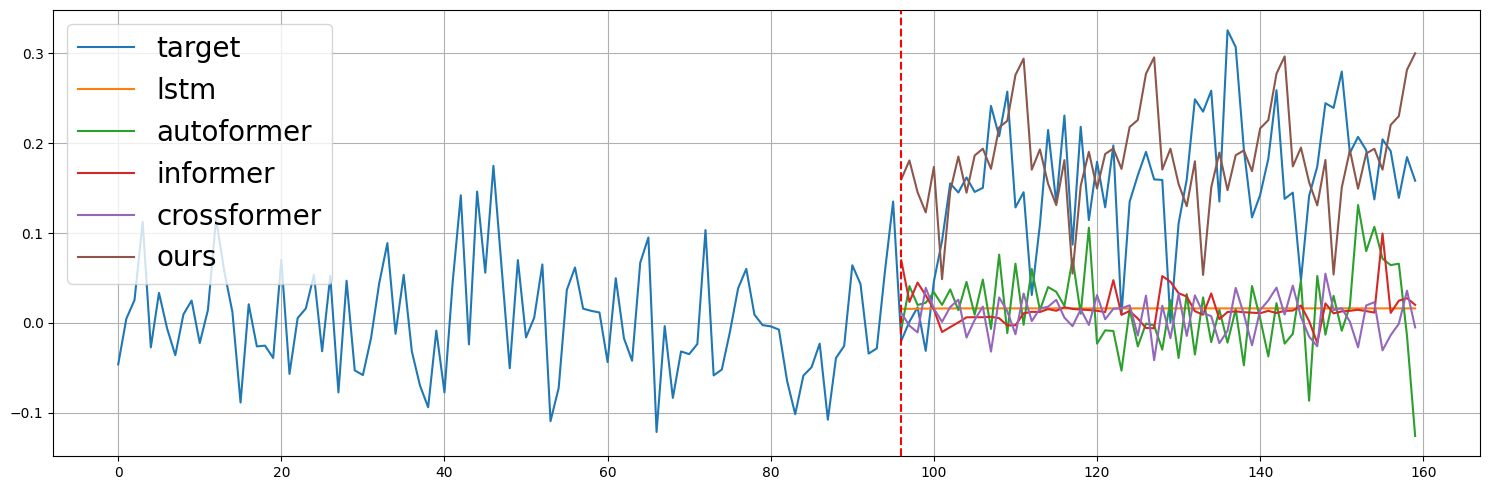}
    \includegraphics[width=0.45\textwidth]{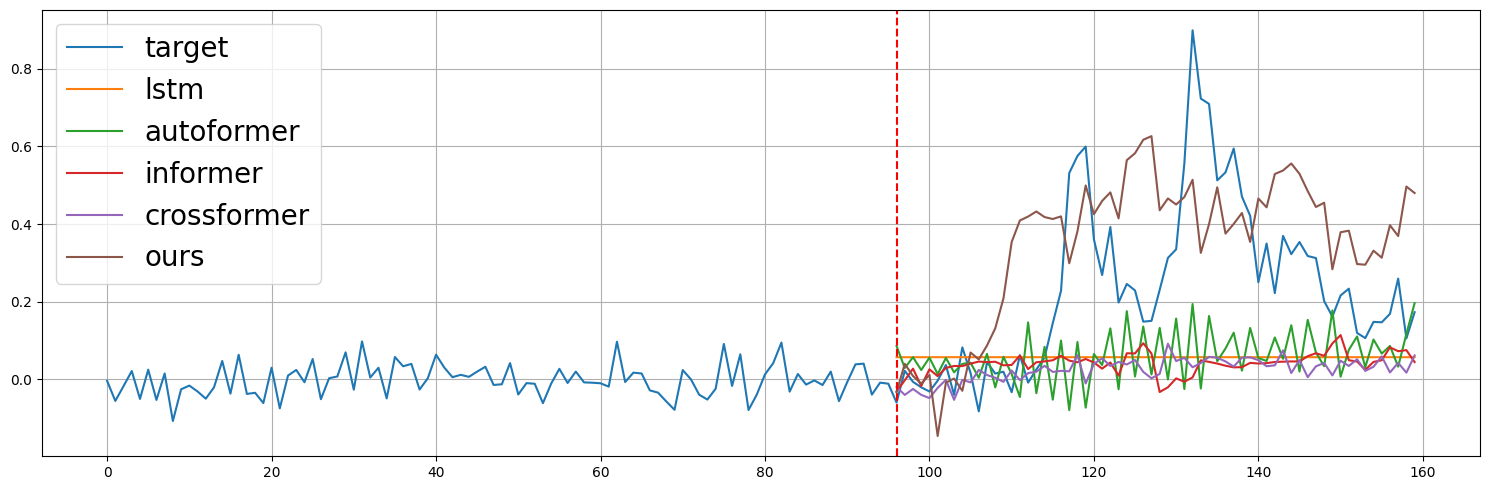}
    \includegraphics[width=0.45\textwidth]{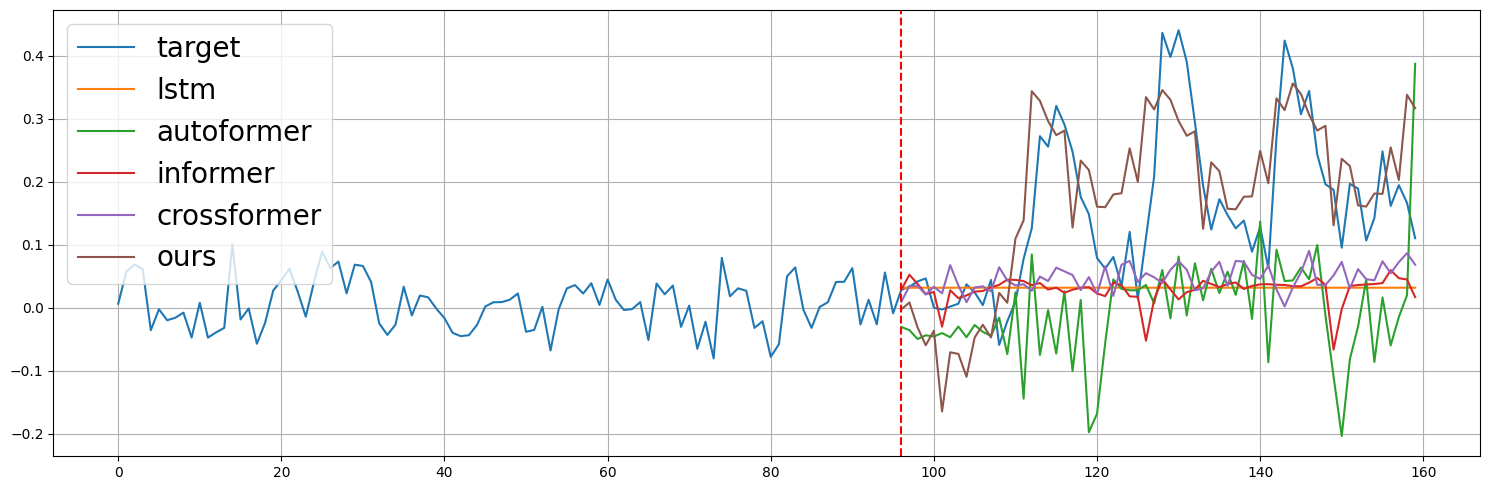}
    \includegraphics[width=0.45\textwidth]{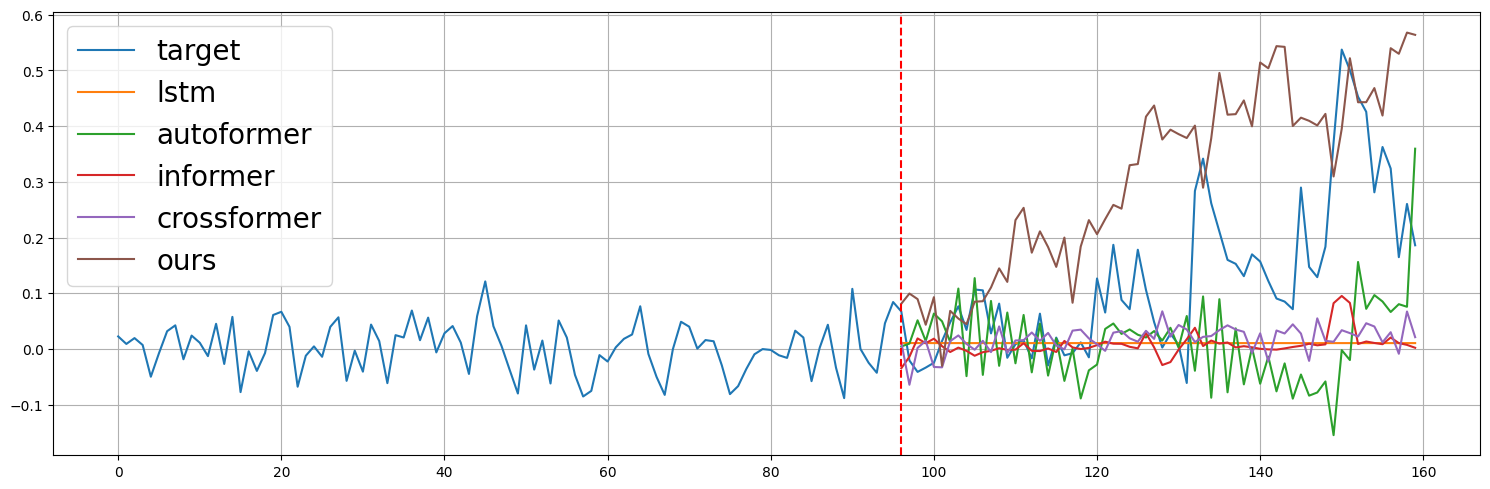}
    \includegraphics[width=0.45\textwidth]{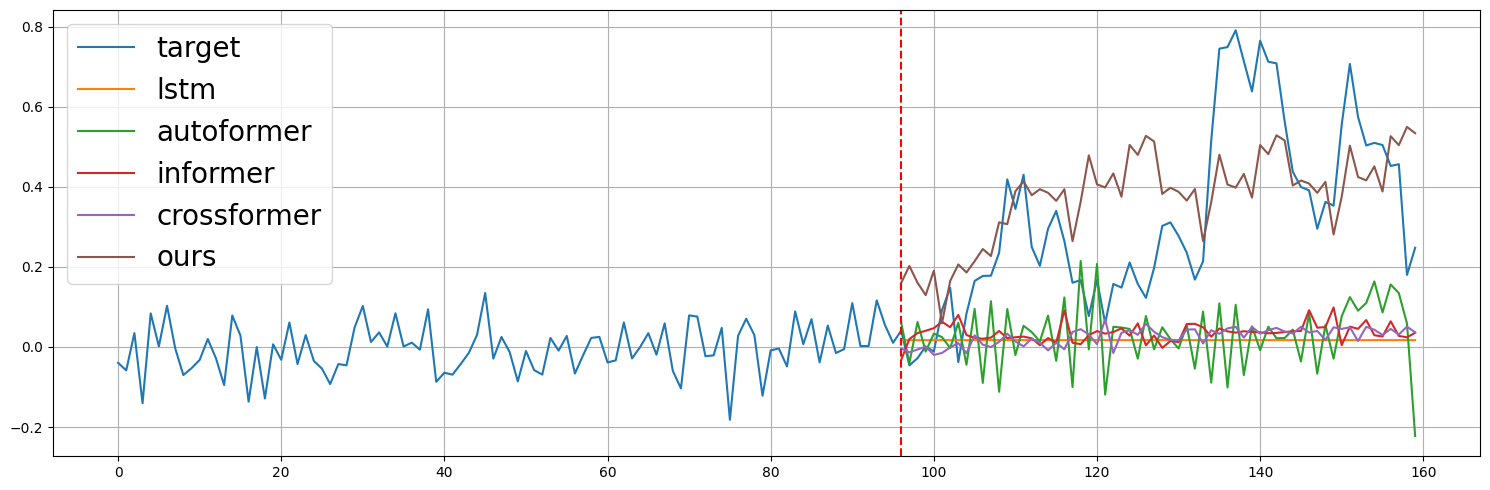}
    
    \caption{\textbf{ Examples of response forecasting by \ours and its competitors on drifting gratings}.}
    \label{supp_fig:drifting}
\end{figure*}

\begin{figure*}[h]
    \centering
    \includegraphics[width=0.45\textwidth]{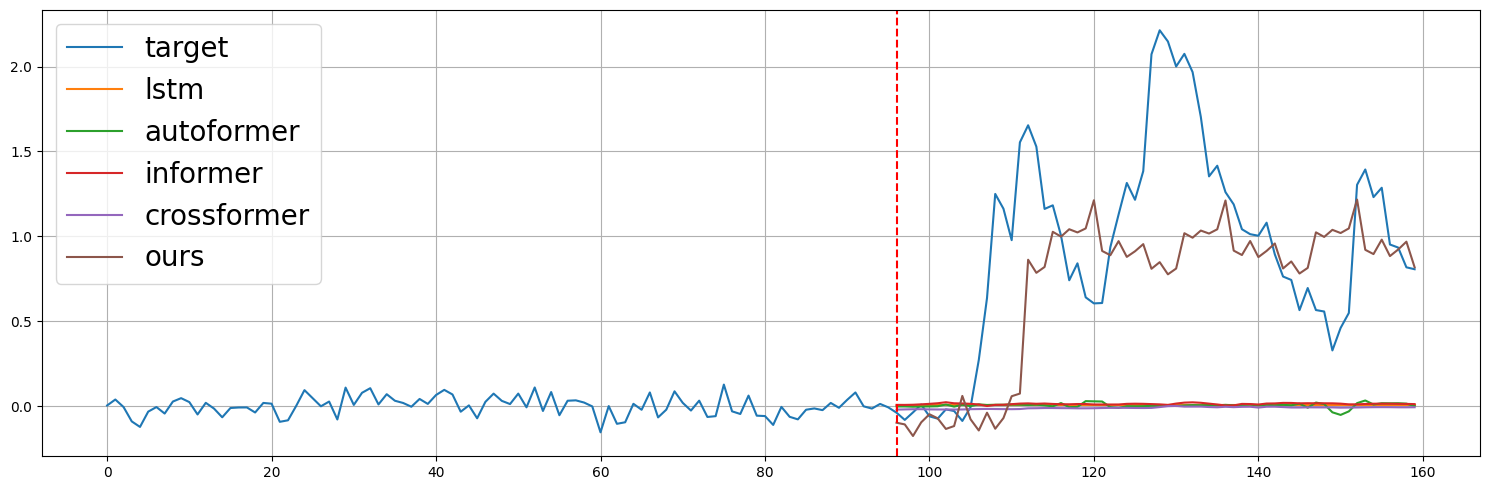}
    \includegraphics[width=0.45\textwidth]{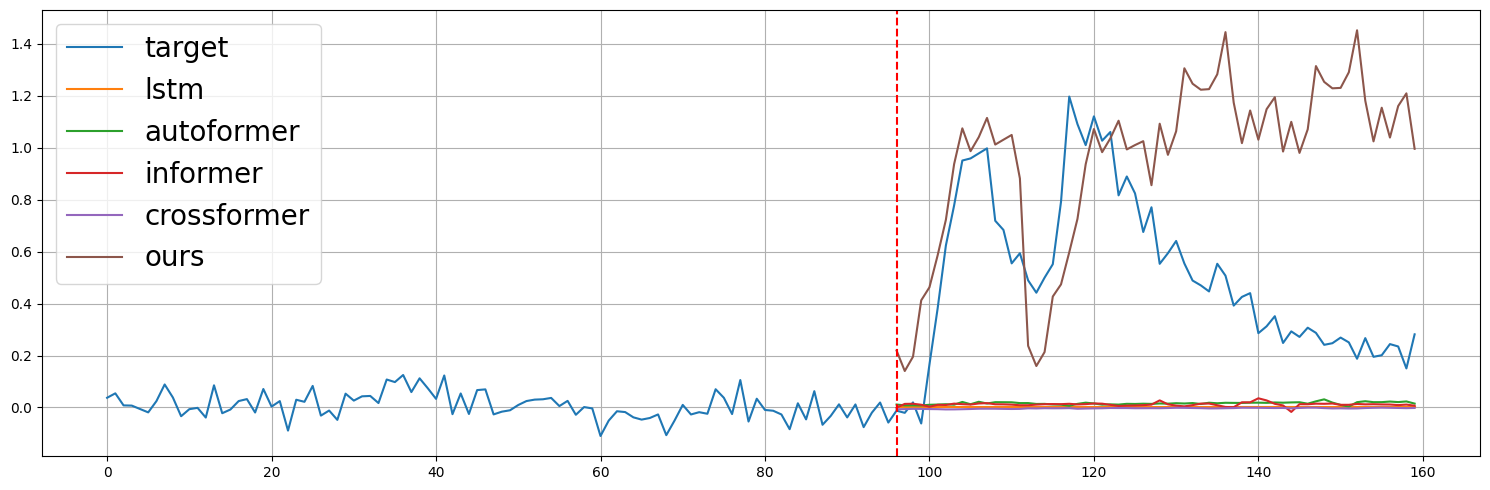}
    \includegraphics[width=0.45\textwidth]{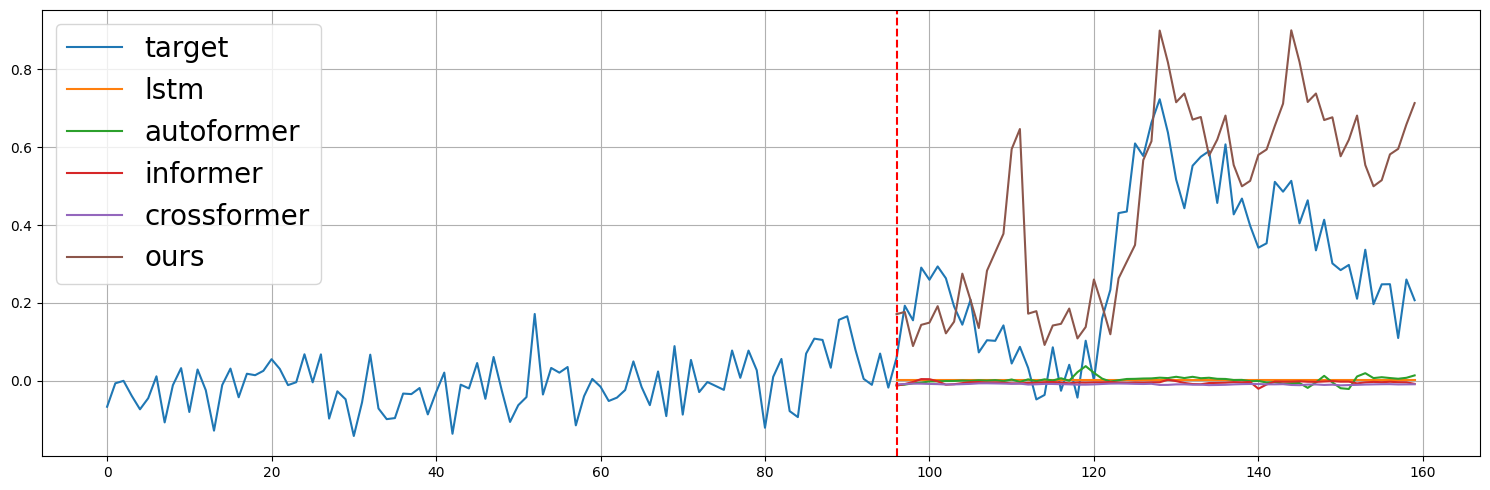}
    \includegraphics[width=0.45\textwidth]{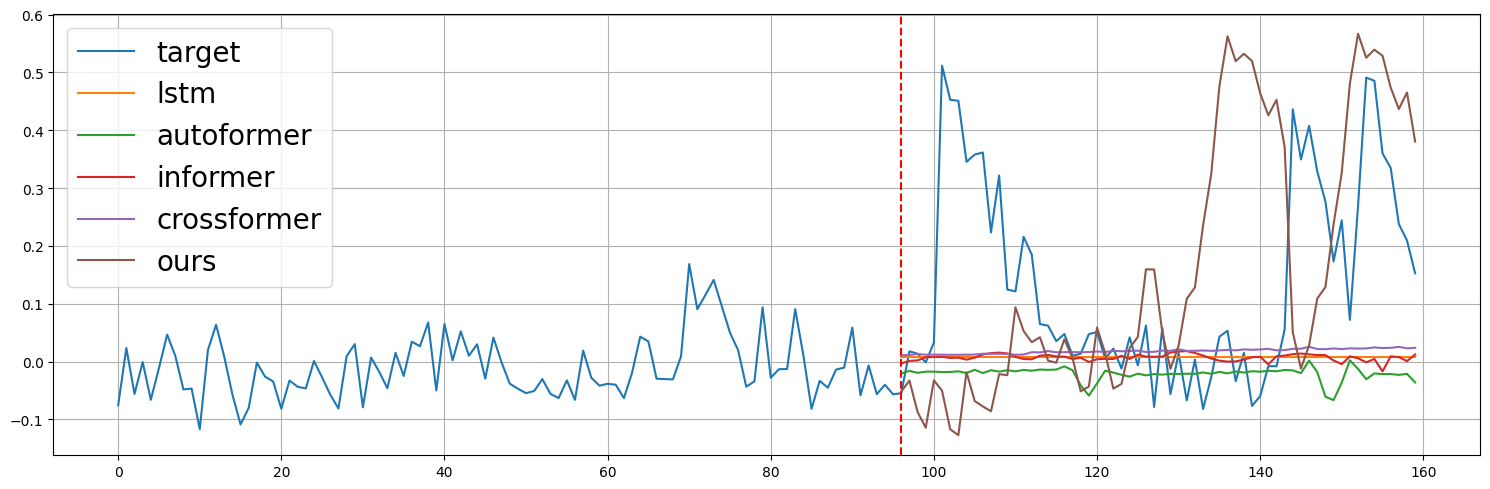}
    \includegraphics[width=0.45\textwidth]{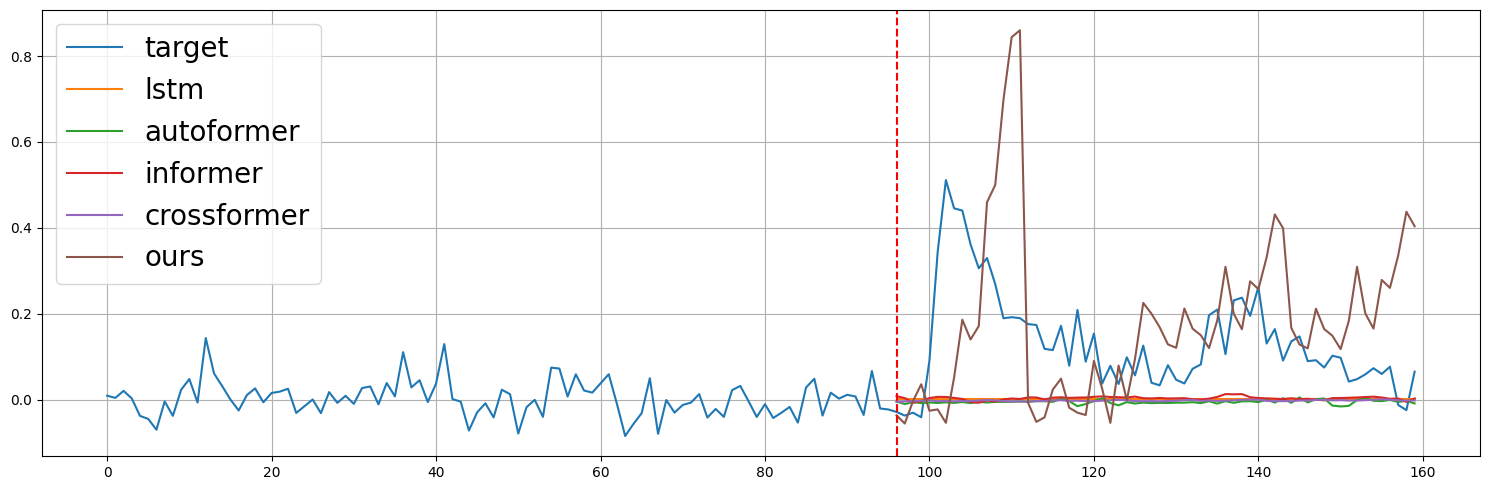}
    \includegraphics[width=0.45\textwidth]{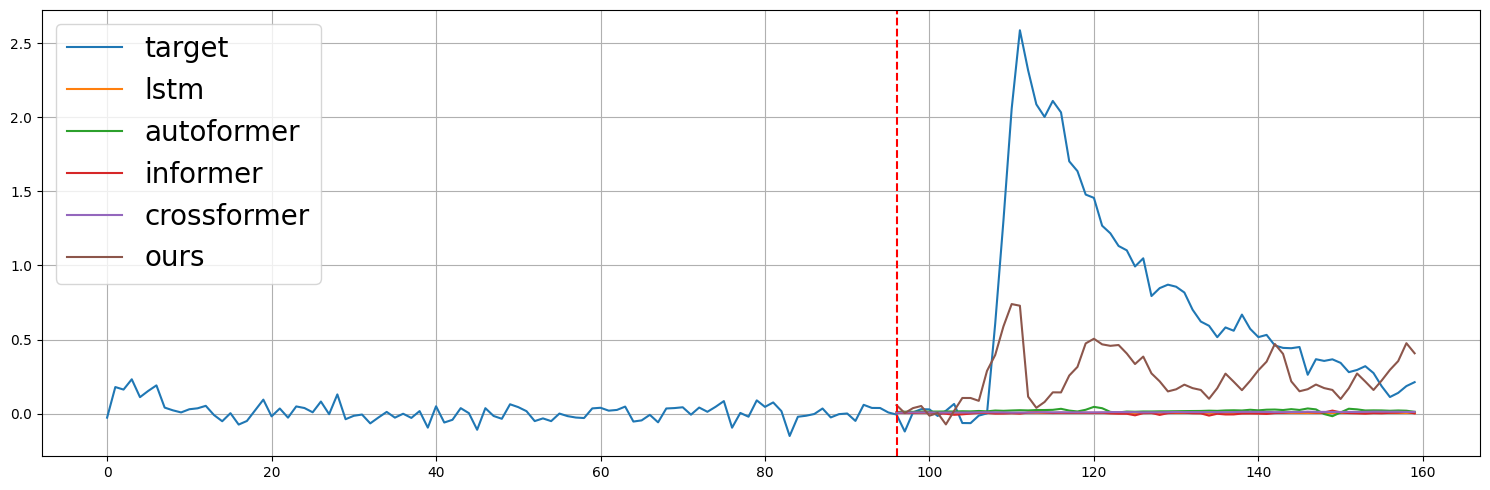}
    
    \caption{\textbf{Examples of response forecasting by \ours and its competitors on static gratings}.}
    \label{supp_fig:static}
\end{figure*}

\begin{figure*}[h]
    \centering
    \includegraphics[width=0.45\textwidth]{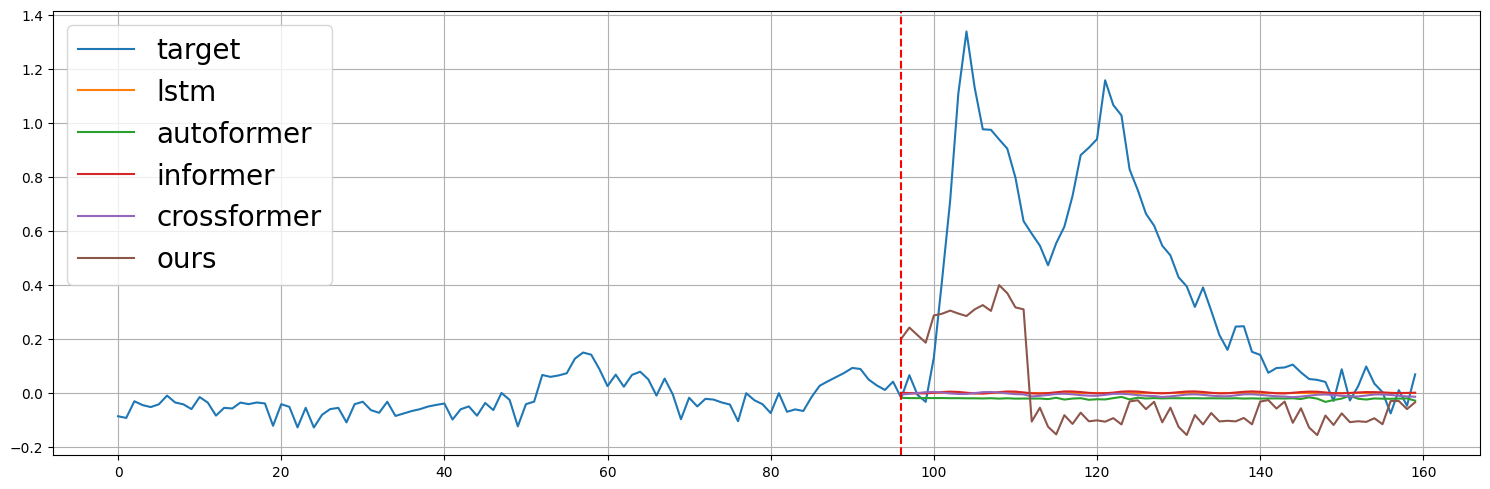}
    \includegraphics[width=0.45\textwidth]{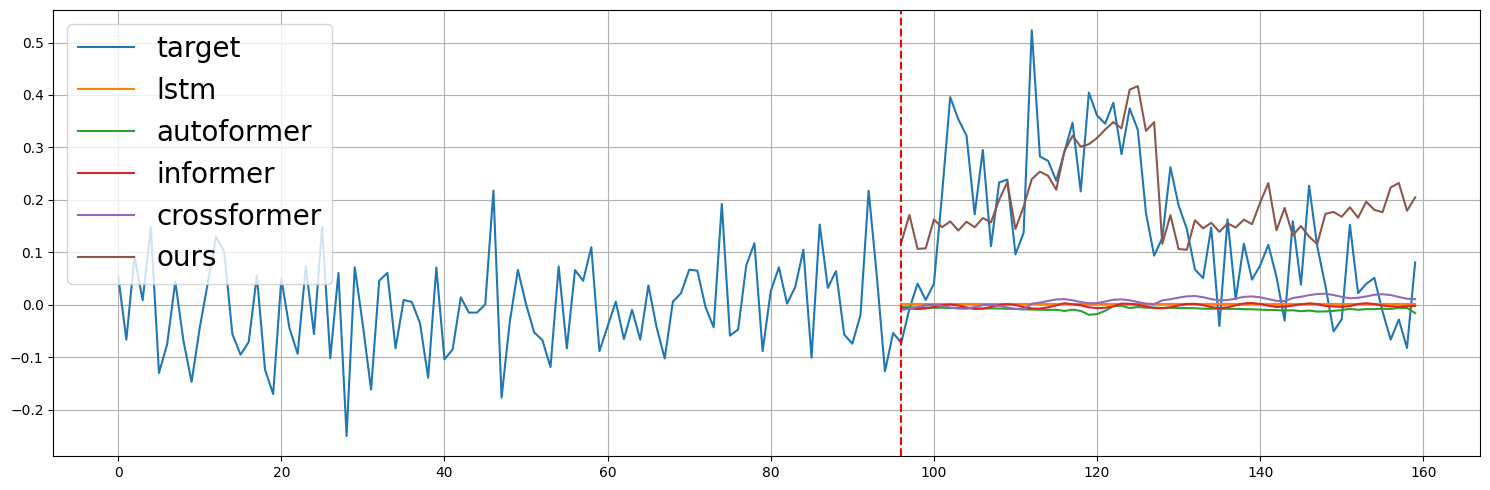}
    \includegraphics[width=0.45\textwidth]{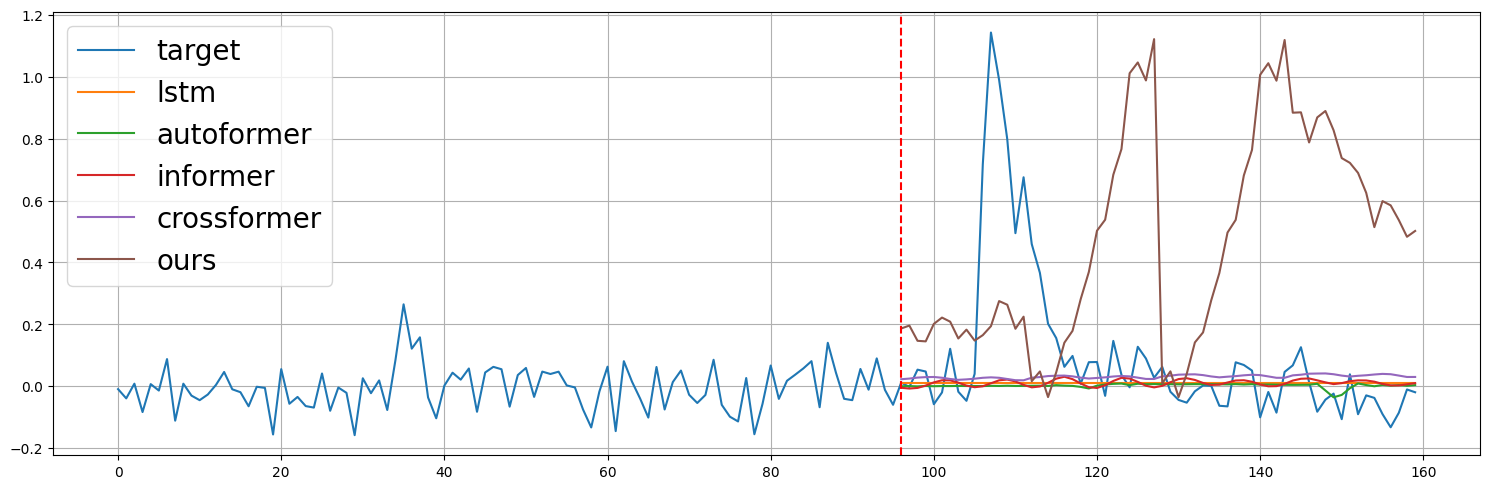}
    \includegraphics[width=0.45\textwidth]{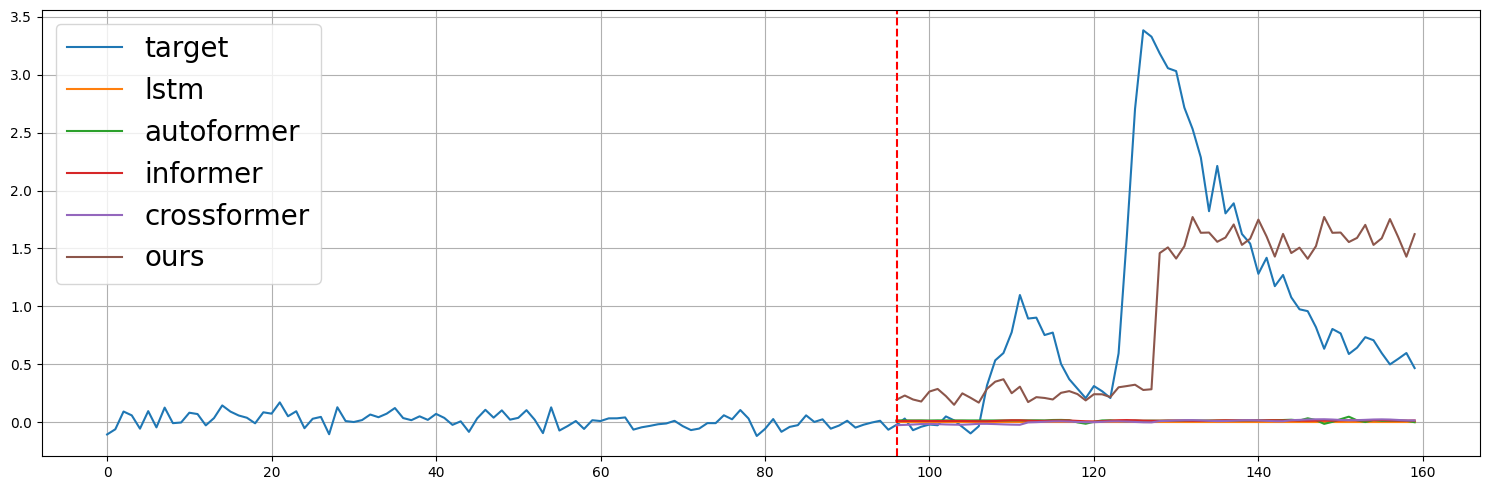}
    \includegraphics[width=0.45\textwidth]{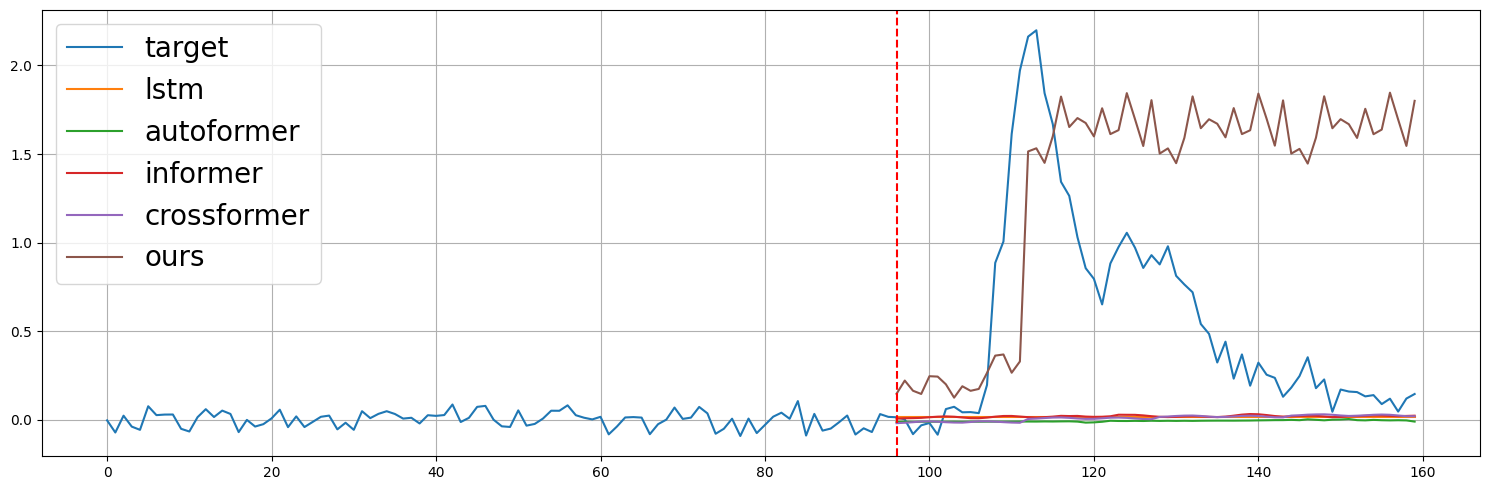}
    \includegraphics[width=0.45\textwidth]{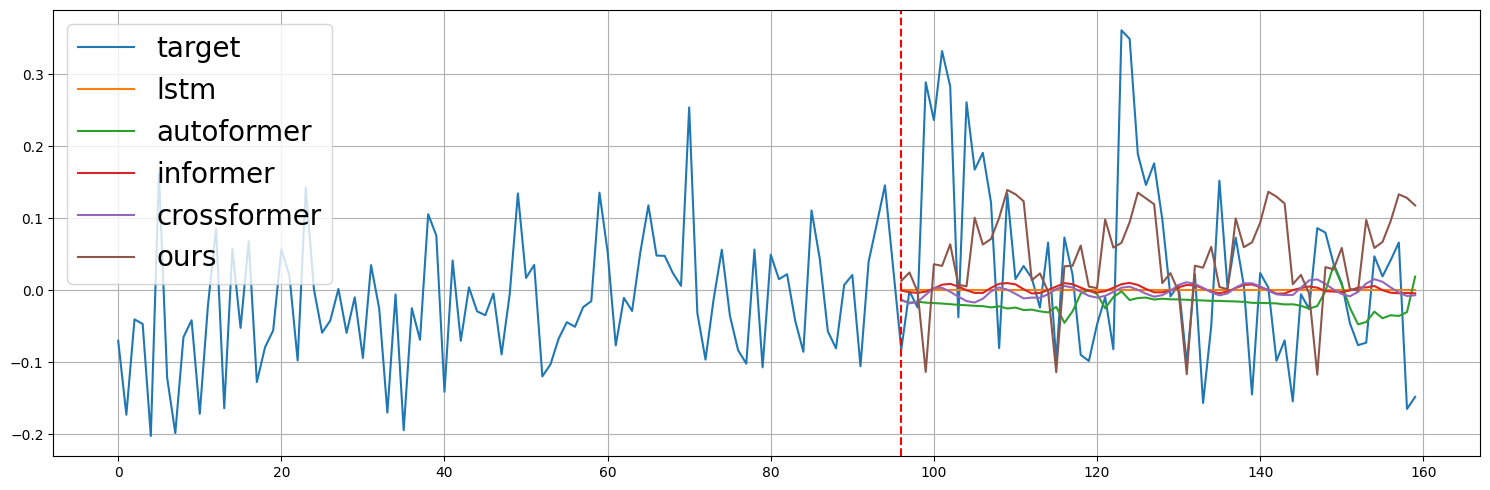}
    
    \caption{\textbf{Examples of response forecasting by \ours and its competitors on locally sparse noise}.}
    \label{supp_fig:noise}
\end{figure*}

\subsection{Application of self-supervised quantization on competitors}
\label{suppl:pretrainig_other}

One might question why our pretraining and quantization strategy was not applied to other methods, especially those based on transformer architectures. The primary reason lies in the substantial modifications required to integrate autoencoding pretraining and quantization into these approaches.

Firstly, quantization is infeasible for Informer~\cite{zhou2021informer} and Autoformer~\cite{autoformer}, due to their reliance on embedding layers along the channel dimension, whereas our method embeds temporally patched data. The goal of quantization is to derive robust temporal representations and patterns. Encoding channel combinations with single codes would create an information bottleneck, emphasizing channel patterns over temporal ones.

Secondly, quantization cannot be directly applied to Crossformer~\cite{zhang2022crossformer}. Although Crossformer performs patching and embedding both channel-wise and temporally, it introduces a two-stage attention mechanism across time and channels. Theoretically, quantization could be implemented; however, pretraining constraints prevent shuffling, altering, or discarding channels. With only 10\% of neurons active per trial, this causes an imbalance during pretraining, leading the quantizer to optimize losses using a limited number of samples for the actual activation. This results in a limited number of quantization codes (3) that mostly describe normal activity signals (the majority in the training data), thus leading to a high quantization error, as shown in Fig.~\ref{fig:crossformer-issue}. Our approach mitigates this by allowing the exclusion of non-active neurons to maintain data balance during pretraining, thus obtaining a much lower quantization error, as shown in Fig.~\ref{fig:crossformer-issue1}.

\begin{figure*}[h!]
    \centering
    \includegraphics[width=0.8\textwidth]{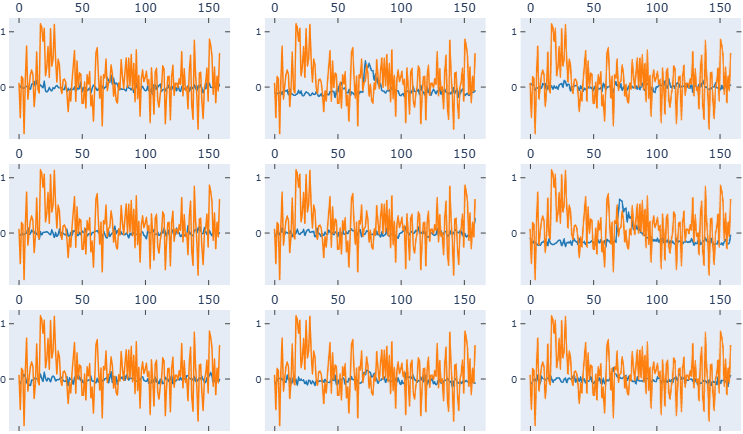}
    \caption{\textbf{Crossformer quantization failure.} In blue, the target signals, while in orange the predicted responses  when using quantization.}
    \label{fig:crossformer-issue}
\end{figure*}

\begin{figure*}[h!]
    \centering
    \includegraphics[width=0.8\textwidth]{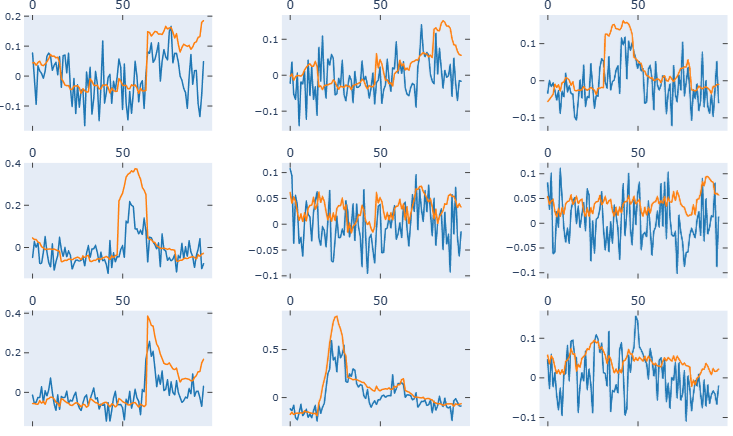}
    \caption{\textbf{\ours quantization performance.} In blue, the target signals, while in orange the predicted responses when using quantization.}
    \label{fig:crossformer-issue1}
\end{figure*}

Furthermore, pretraining itself is problematic for similar reasons. Different containers possess unique channels, necessitating significant alterations to existing methods for effective pretraining. For Autoformer and Informer, each container and experiment would require a dedicated embedding layer to map input channel dimensions into a unified latent space. For Crossformer, introducing a pad token and padding mask might make pretraining feasible, but there would be no consistency in channel order across different containers and experiments. This inconsistency would result in channel attention learning non-generalizable dependencies. Even within a single container, such as a mouse, the number of channels and their order vary across experiments.

Thus, our strategy is more appropriate for pretraining, given the inherent challenges and limitations of adapting other methods for this purpose.

\qftext{
\subsection{Post-hoc deconvolution consistency analysis}

\qftext{As an additional robustness check, we evaluated whether fluorescence forecasts preserve information after spike deconvolution. Since all models are trained to predict fluorescence traces rather than supervised spike/event labels, this analysis is intended as a post-hoc consistency test rather than as a direct spike-forecasting benchmark. We applied the same normalization and OASIS \cite{friedrich2017fast} deconvolution pipeline to both predicted and target fluorescence traces, and report continuous deconvolved-rate metrics together with thresholded-event balanced accuracy as a diagnostic.

As shown in Table~\ref{tab:deconv}, the thresholded-event balanced accuracy is close to chance for all methods, indicating that post-hoc event recovery is highly sensitive to deconvolution scale and threshold selection. Overall, this analysis confirms that our main conclusions should be interpreted at the level of fluorescence forecasting, while direct supervised spike/event forecasting remains an important direction for future work.}
}

\begin{table*}[htb!]
    \centering
    \renewcommand{\arraystretch}{1.2}
    \caption{\textbf{Post-hoc deconvolved-rate consistency analysis on drifting gratings stimuli.}}
    \label{tab:deconv}
    \footnotesize{
    \begin{tabular}{l|c|ccc}
        \toprule
        & \multicolumn{1}{c|}{\textbf{Thresholded events}} & \multicolumn{3}{c}{\textbf{Deconvolved-rate Forecasting}}\\
        \midrule
        \textbf{Method} & \textbf{Bal. Acc.} ($\uparrow$) & \textbf{Corr.} ($\uparrow$) & \textbf{MAE} ($\downarrow$) & \textbf{MSE} ($\downarrow$)\\
        \midrule
        \rowcolor{gray!10}
        LSTM & \resultnof{50.57}{0.10} & \resultnof{0.0040}{0.0012} & \resultnof{0.0211}{0.0034} & \resultnof{0.0017}{0.0006} \\
        Autoformer & \resultnof{50.43}{0.10} & \resultnof{0.0028}{0.0023} & \resultnof{0.0085}{0.0008} & \resultnof{0.0004}{0.0000} \\
        \rowcolor{gray!10}
        Informer & \resultnof{50.82}{0.15} & \resultnof{0.0046}{0.0032} & \resultnof{0.0084}{0.0012} & \resultnof{0.0003}{0.0001} \\
        BrainLM & \resultnof{50.90}{1.00} & \resultnof{{\textbf{0.0111}}}{0.0093} & \resultnof{0.0115}{0.0056} & \resultnof{0.0005}{0.0003} \\
        \rowcolor{gray!10}
        BrainLM$_{ft}$ & \resultnof{50.85}{0.60} & \resultnof{0.0099}{0.0074} & \resultnof{0.0090}{0.0031} & \resultnof{0.0003}{0.0001} \\
        Crossformer & \resultnof{50.66}{0.15} & \resultnof{0.0093}{0.0045} & \resultnof{0.0070}{0.0004} & \resultnof{0.0002}{0.0000} \\
        \rowcolor{gray!10}
        \ours & \resultnof{51.10}{0.56} & \resultnof{0.0090}{0.0087} & \resultnof{0.0088}{0.0007} & \resultnof{0.0003}{0.0000} \\
        \bottomrule
    \end{tabular}}
    \label{tab:deconvolution_forecasting_drifting}
\end{table*}

\subsection{Interpretability of learned codes}
\label{suppl:learned_codes}

Fig.~\ref{fig:codes} shows the latent space structure of discrete codes learned by vector quantization. We performed 2D t-SNE on a learned codebook to observe sequence patterns. Subfigure (a) shows that amplitude increases along the x-axis when plotting codes on the same scale. Subfigure (b) reveals pattern variability after normalizing the scale. Interestingly, despite having a relatively small number of codes, the reconstructed representation heavily depends on the sequence, as shown in Subfigure (c): we generated sequences with bursts of the same code, except for one typically representing a peak (e.g., code 19), highlighted between red dashed lines. The replaced code's amplitude and shape vary based on context, indicating that while codes represent patterns, the reconstruction depends on the whole sequence. 

\begin{figure*}[!ht]
    \centering
    \includegraphics[width=0.42\textwidth]{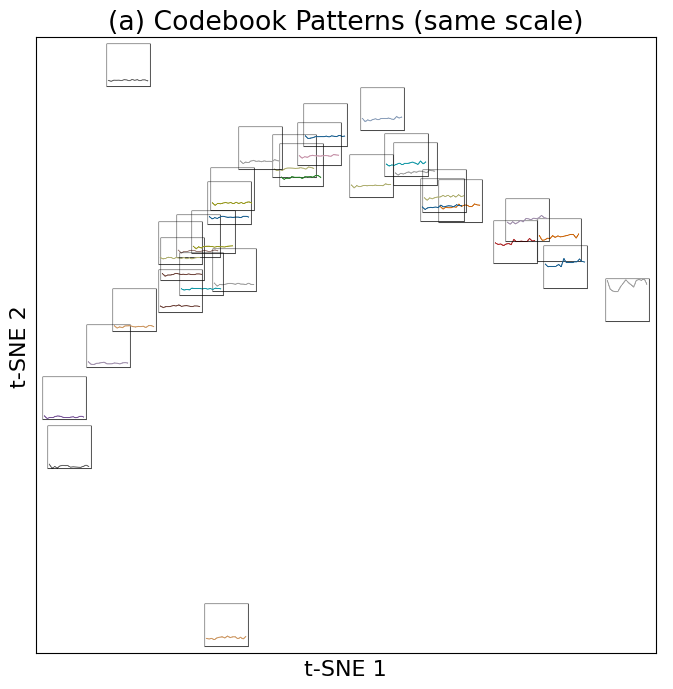}
    \includegraphics[width=0.42\textwidth]{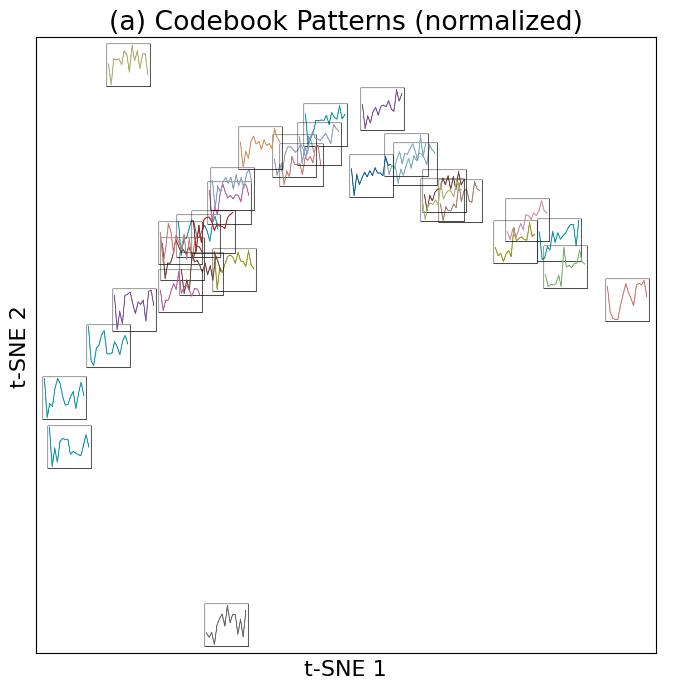}
    \includegraphics[width=0.85\textwidth]{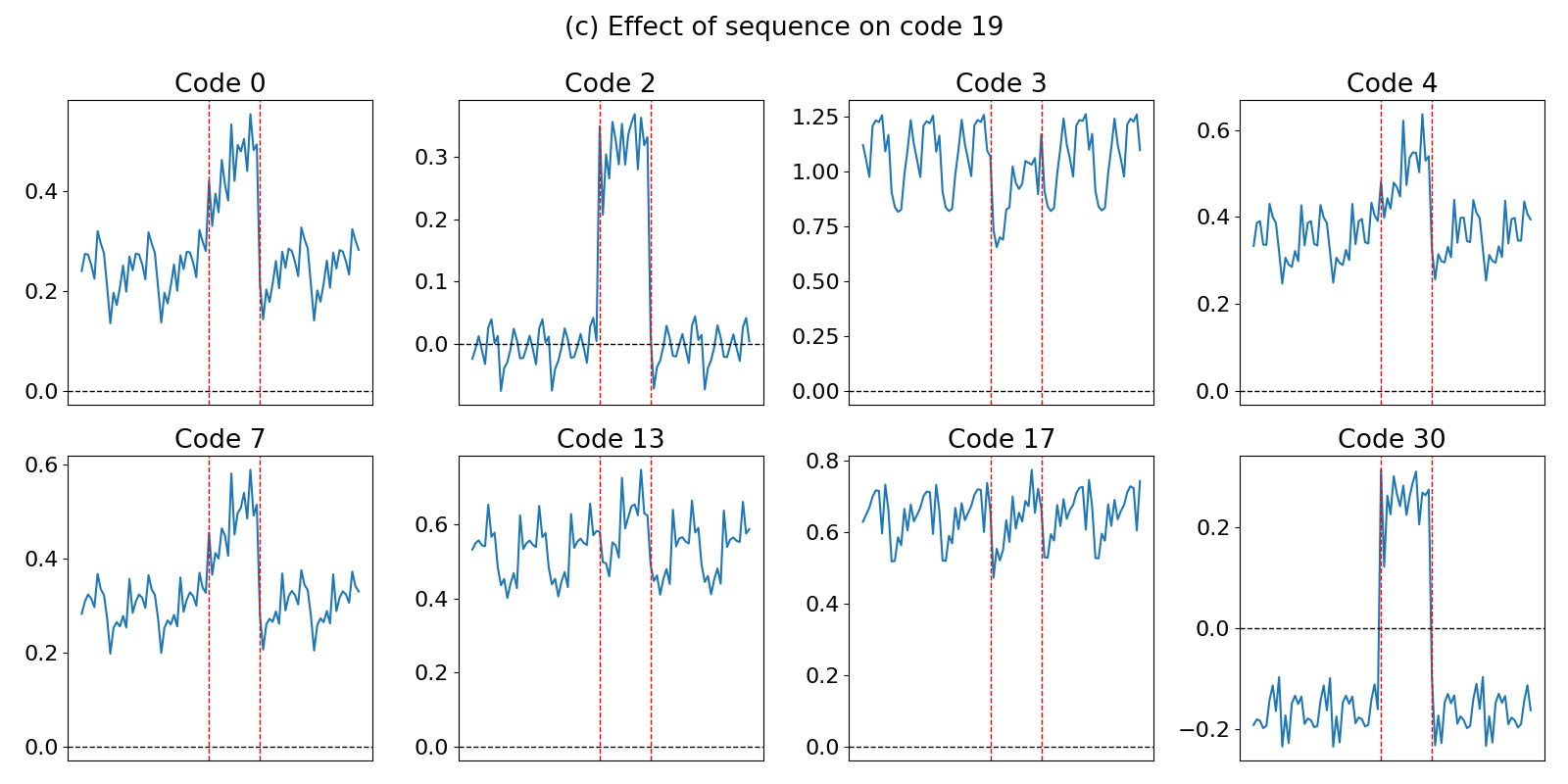}
    \caption{Interpretability of codes. (a) t-SNE of a codebook, with patterns representation in the same scale. The first axis captures amplitude variation. (b) Same as before, but with normalization to appreciate differences in patterns. (c) Effect of sequence.}
    \label{fig:codes}
\end{figure*}

\subsection{Interpretability of neuron embeddings}
\label{sec:neuron_embs}

To \qftext{understand} what is encoded into neuron embeddings, we visualized through t-SNE neuron embeddings from a downstream task. We find that neuron embeddings encode information such as activation frequency and response statistics. Colors in Fig.~\ref{fig:neuron_embs} denote whether the measured quantity is above or below a threshold.

\begin{figure*}[!ht]
    \centering
    \includegraphics[width=0.90\textwidth]{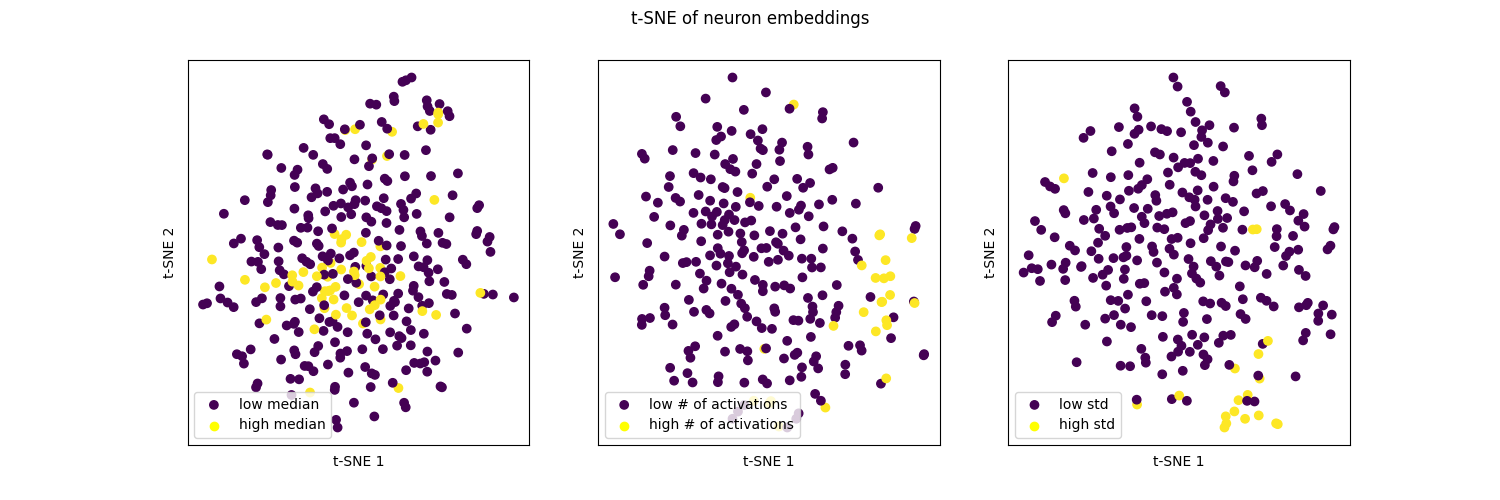}
    \caption{Interpretability of neuron embeddings. We show t-SNE examples of neuron embeddings. We found that similar neurons in the latent space have also similar statistics like the median, the number of activations or the standard deviation.}
    \label{fig:neuron_embs}
\end{figure*}

\endgroup

\clearpage
\bibliographystyle{unsrtnat}
\bibliography{bibliography}

\end{document}